\documentclass[11pt,english,a4paper]{article}
\usepackage[T1]{fontenc}
\usepackage[latin1]{inputenc}
\usepackage{amsmath}
\usepackage{graphicx}
\usepackage{amssymb}
\usepackage{amsfonts}
\usepackage{mathrsfs}
\usepackage{epsfig}

\newcounter{algo}
\newenvironment{algo}[2]{\refstepcounter{algo}\label{#2}   \begin{center}
\begin{minipage}{0.9\textwidth}   \hrule\smallskip
\textbf{Algorithm \thealgo: #1}
\par\smallskip\hrule\smallskip\ignorespaces}{\par\smallskip\hrule
\end{minipage}
\end{center}
}

\newtheorem{theorem}{Theorem}
\newtheorem{proposition}[theorem]{Proposition}
\newtheorem{definition}[theorem]{Definition}
\newtheorem{lemma}[theorem]{Lemma}
\newtheorem{corollary}[theorem]{Corollary}

\newtheorem{remark}[theorem]{Remark}
\newenvironment{proof}[1][Proof]{\noindent \textbf{#1.} }{\qedsymbol}

\newcommand{\qedsymbol}{\hspace{\fill}\rule{1.5ex}{1.5ex}}
\def\baselinestretch{1.38}
\usepackage{babel}

\begin{document}

\title{\vspace{-2.1cm}
 {\Huge Distributed Power Allocation with Rate Constraints in Gaussian
Parallel Interference Channels}}

\date{{\small Submitted to }\textit{\small IEEE Transactions on Information
Theory}{\small , February 17, 2007.}{\normalsize }\\
{\normalsize{} }{\small {} Revised January 11, 2008. }}

\author{Jong-Shi Pang$^{1,}$%
\thanks{The work of this author is based on research supported by the National
Science Foundation under grant DMI-0516023.%
}\ , Gesualdo Scutari$^{2,\natural,}$%
\thanks{The work of this author is based on research partially supported by
the SURFACE project funded by the European Community under Contract
IST-4-027187-STP-SURFACE, and by the Italian Ministry of University
and Research.%
}\ , Francisco Facchinei$^{3,}$%
\thanks{The work of this author is based on research supported by the Program
PRIN-MIUR 2005 {}``Innovative Problems and Methods in Nonlinear Optimization\char`\"{}.%
}, and Chaoxiong Wang$^{4}$.\\
 {\small Email: }\texttt{\small jspang@uiuc.edu, scutari@infocom.uniroma$1$.it,
facchinei@dis.uniroma$1$.it, wangc@rpi.edu.}{\small \vspace{0.2cm}
 }{\normalsize }\\
{\normalsize{} {} {} }{\small $^{1}$ Dept. of Industrial and Enterprise Systems Engineering,
University of Illinois at Urbana-Champaign,\vspace{-0.2cm}}\\{\small 104 S. Mathews Ave., Urbana IL 61801,  U.S.A..}{\normalsize \vspace{0.2cm}
 }\\
{\normalsize{} {} {} }{\small $^{2}$ Dept. INFOCOM, University of
Rome {}``La Sapienza\char`\"{}, Via Eudossiana 18, 00184, Rome, Italy.}{\normalsize \vspace{0.2cm}
 }\\
{\normalsize{} {} {} }{\small $^{3}$ Dept. of Computer and Systems
Science {}``Antonio Ruberti\char`\"{}, University of Rome {}``La
Sapienza\char`\"{},\vspace{-0.2cm}
 }{\normalsize }\\
{\normalsize{} {} }{\small Via Ariosto 25, 00185 Rome, Italy.}{\normalsize }\\
{\normalsize{} {} {} }{\small $^{4}$ Dept. of Mathematical Sciences,
Rensselaer Polytechnic Institute, Troy, New York, 12180-3590, U.S.A..}{\normalsize \vspace{0.2cm}
 }\\
{\normalsize{} {} {} { $^{\natural}$Corresponding author}}}

\maketitle
\vspace{-0.3cm}

\begin{abstract}
\noindent This paper considers the minimization of transmit power
in Gaussian parallel interference channels, subject to a rate constraint
for each user. To derive decentralized solutions that do not require
any cooperation among the users, we formulate this power control problem
as a (generalized) Nash equilibrium game. We obtain sufficient conditions
that guarantee the existence and nonemptiness of the solution set
to our problem. Then, to compute the solutions of the game, we propose
two distributed algorithms based on the single user waterfilling solution:
The \emph{sequential} and the \emph{simultaneous} iterative waterfilling
algorithms, wherein the users update their own strategies sequentially
and simultaneously, respectively. We derive a unified set of sufficient
conditions that guarantee the uniqueness of the solution and global
convergence of both algorithms. Our results are applicable to all
practical distributed multipoint-to-multipoint interference systems,
either wired or wireless, where a quality of service in terms of information
rate must be guaranteed for each link.

\bigskip{}
\bigskip{}

\vfill

\end{abstract}

\newpage{}

\section{Introduction and Motivation}

The interference channel is a mathematical model relevant to many
communication systems where multiple uncordinated links share a common
communication medium, such as wireless ad-hoc networks or Digital
Subscriber Lines (DSL). In this paper, we focus on the Gaussian parallel
interference channel. 

A pragmatic approach that leads to an achievable region or inner bound
of the capacity region is to restrict the system to operate as a set
of independent units, i.e., not allowing multiuser encoding/decoding
or the use of interference cancelation techniques. This achievable
region is very relevant in practical systems with limitations on the
decoder complexity and simplicity of the system. With this assumption,
multiuser interference is treated as noise and the transmission strategy
for each user is simply his power allocation. The system design reduces
then to finding the optimum power distribution for each user over
the parallel channels, according to a specified performance metric.

Within this context, existing works \cite{Cendrillon-Yu}-\cite{Scutari-Palomar-Barbarossa-IT_Journal}
considered the maximization of the information rates of all the links,
subject to transmit power and (possibly) mask constraints on each
link. In \cite{Cendrillon-Yu}-\cite{Yu-Lui}, a centralized approach
based on duality theory \cite{Bazaraa-Sherali-Shetty,Boyd} was proposed
to compute, under technical conditions, the largest achievable rate
region of the system (i.e., the Pareto-optimal set of the achievable
rates). 
In \cite{Hayashi-Luo}, sufficient conditions for the optimal spectrum
sharing strategy maximizing the sum-rate to be frequency division
multiple access (FDMA) were derived. However, the algorithms proposed in \cite{Cendrillon-Yu}-\cite{Hayashi-Luo} are computationally expensive and cannot be implemented in a distributed way, require the full-knowledge of the system parameters, and are not guaranteed to converge to the global optimal solution. 

Therefore, in \cite{Yu}-\cite{Scutari-Palomar-Barbarossa-IT_Journal}, using a game-theory framework, 
the authors focused on distributed algorithms with no centralized
control. In particular, the rate maximization problem
was formulated as a strategic non-cooperative game, where every link
is a player that competes against the others by choosing the transmission
strategy that maximizes its own information rate \cite{Yu}. 
Based on the celebrated
notion of Nash Equilibrium (NE) in game theory \cite{Nash-paper},
an equilibrium for the whole system is reached when every player's
reaction is ``unilaterally optimal'',  i.e., when, given the rival
players' current strategies, any change in a player's own strategy
would result in a rate loss. In \cite{ChungISIT03}-\cite{Scutari-Palomar-Barbarossa-IT_Journal},
alternative sufficient conditions were derived that guarantee the
uniqueness of the NE of the rate maximization game and the convergence
of alternative distributed waterfilling based algorithms, either synchronous
$-$ sequential \cite{ChungISIT03}-\cite{LPang06} and simultaneous
\cite{Scutari-Barbarossa-GT_PII} $-$ or asynchronous \cite{Scutari-Palomar-Barbarossa-IT_Journal}.

The game theoretical formulation proposed in the cited papers, is
a useful approach to devise totally distributed algorithms. However,
due to possible asymmetries of the system and the inherent selfish
nature of the optimization, the Nash equilibria of the rate maximization
game in \cite{Yu}-\cite{Scutari-Palomar-Barbarossa-IT_Journal} may
lead to inefficient and unfair rate distributions among the links 
even when the game admits a unique NE. This unfairness is due to the
fact that, without any additional constraint, the optimal power allocation
corresponding to a NE of the rate maximization game  is often
the one that assigns high rates to the users with the highest (equivalent)
channels; which strongly penalizes all the other users. As many realistic
communication systems require prescribed Quality of Service (QoS)
guarantees in terms of achievable rate for each user, the system design
based on the game theoretic formulation of the rate maximization might
not be adequate.

To overcome this problem, in this paper we introduce a new distributed
system design, that takes explicitly into account the rate constraints.
More specifically, we propose a novel strategic non-cooperative game,
where every link is a player that competes against the others by choosing
the power allocation over the parallel channels that attains the desired
information rate, with the minimum transmit power. We will refer to
this new game as \emph{power minimization game}. An equilibrium is
achieved when every user realizes that, given the current power allocation
of the others, any change in its own strategy would result in an increase
in transmit power. This equilibrium is referred to as \emph{Generalized}
Nash Equilibrium (GNE) and the corresponding game is called Generalized
Nash Equilibrium Problem.%
\footnote{According to recent use, we term generalized Nash equilibrium problem
a Nash game where the feasible sets of the players depend on the other
players' strategy. Such kind of games have been called in various
different ways in the literature, for example social equilibrium problems
or just Nash equilibrium problems.}

The game theoretical formulation proposed in this paper differs significantly
from the rate maximization games studied in \cite{Yu}-\cite{Scutari-Palomar-Barbarossa-IT_Journal}.
In fact, differently from these references, where the users are allowed
to choose their own strategies independently from each other, in the
power minimization game, the rate constraints induce a coupling among
the players' admissible strategies, i.e., each player's strategy set
depends on the current strategies of all the other players. This coupling
makes the study of the proposed game much harder than that of the
rate maximization game and no previous result in \cite{Yu}-\cite{Scutari-Palomar-Barbarossa-IT_Journal}
can be used. Recently, the calculation of generalized Nash equilibria
has been the subject of a renewed attention also in the mathematical
programming community, see for example \cite{FPang03}-\cite{Facchinei-Kanzow}.
Nevertheless, in spite of several interesting advances \cite{Facchinei-Kanzow}, none of the
game results in the literature are applicable to the power minimization
game.

The main contributions of the paper are the following. We provide
sufficient conditions for the nonemptiness and boundedness of the
solution set of the generalized Nash problem. Interestingly, these
sufficient conditions suggest a simple admission control procedure
to guarantee the feasibility of a given rate profile of the users. Indeed, our existence proof uses an advanced degree-theoretic result
for a nonlinear complementarity problem in order to handle the
unboundedness of the users' rate constraints.
We also derive conditions for the uniqueness of the GNE.  
Interestingly, our sufficient conditions become also necessary in the case of one subchannel. 
To compute
the generalized Nash solutions, we propose two alternative totally
distributed algorithms based on the single user waterfilling solution:
The \emph{sequential} IWFA and the \emph{simultaneous} IWFA. The sequential
IWFA is an instance of the Gauss-Seidel scheme: The users update their
own strategy sequentially, one after the other, according to the single
user waterfilling solution and treating the interference generated
by the others as additive noise. The simultaneous IWFA is based on
the Jacobi scheme: The users choose their own power allocation simultaneously,
still using the single user waterfilling solution. Interestingly,
even though the rate constraints induce a coupling among the feasible
strategies of all the users, both algorithms are still totally distributed.
In fact, each user, to compute the waterfilling solution, only needs
to measure the power of the noise plus the interference generated by the other users over each subchannel.  It
turns out that the conditions for the uniqueness of the GNE are sufficient
for the convergence of both algorithms. Our convergence
analysis is based on a nonlinear transformation that turns the
generalized game in the power variables into a standard game in the
rate variables.  Overall, this paper offers two major contributions
to the literature of game-theoretic approaches to multiuser
communication systems: (i) a new noncooperative game model is
introduced for the first time that directly addresses the issue of
QoS in such systems, and (ii) a new line of analysis
is introduced in the literature of distributed power allocation that
is expected to be broadly applicable for other game models.

The paper is organized as follows. Section \ref{sec:systMod_GTformulation}
gives the system model and formulates the power minimization problem
as a strategic non-cooperative game. Section \ref{Sec:Existence_Uniqueness}
provides the sufficient conditions for the existence and uniqueness
of a GNE of the power minimization game. Section \ref{Sec:IWFAs}
contains the description of the distributed algorithms along with
their convergence conditions. Finally, Section \ref{Sec:Conclusions}
draws the conclusions. Proofs of the results are given in the Appendices
\ref{proof_th:existence(main_body)}--\ref{proof_th:Convergence_IWFA-SIWFA}.

\section{System Model and Problem Formulation}

\label{sec:systMod_GTformulation}In this section we clarify the assumptions
and the constraints underlying the system model and we formulate the
optimization problem explicitly.

\subsection{System model}

\label{Sec:System.Model} We consider a $Q$-user Gaussian $N$-parallel
interference channel. In this model, there are $Q$ transmitter-receiver
pairs, where each transmitter wants to communicate with its corresponding
receiver over a set of $N$ parallel subchannels. These subchannels
can model either frequency-selective or flat-fading time-selective
channels \cite{Tse}. 
Since our goal is to find distributed algorithms that do not require
neither a centralized control nor a coordination among the links, we
focus on transmission techniques where no interference cancelation
is performed and multiuser interference is treated as additive colored
noise from each receiver. Moreover, we assume perfect channel state
information at both transmitter and receiver side of each link;\footnote{Note that each user is only required to known its own channel, but not the channels of the other users.} each
receiver is also assumed to measure with no errors the power of the
noise plus the overall interference generated by the other users over
the $N$ subchannels. For each transmitter $q$, the total average
transmit power over the $N$ subchannels is (in units of energy per
transmitted symbol) \begin{equation}
P_{q}=\dfrac{1}{N}\sum_{k=1}^{N}p_{q}(k),\label{Tx-power}\end{equation}
 where $p_{q}(k)$ denotes the power allocated by user $q$ over the
subchannel $k$.

Under these assumptions, invoking the capacity expression for the
single user Gaussian channel $-$ achievable using random Gaussian
codes from all the users $-$ the maximum information rate on link
$q$ for a specific power allocation is \cite{Cover}%
\footnote{Observe that a GNE is obtained if each user transmits using Gaussian
signaling, with a proper power allocation. However, generalized Nash
equilibria achievable using non-Gaussian codes may exist. In this paper,
we focus only on transmissions using Gaussian codebooks.%
} \begin{equation}
R_{q}(\mathbf{p}_{q},\mathbf{p}_{-q})=\sum_{k=1}^{N}\log\left(1+\mathsf{sinr}_{q}(k)\right),\label{Rate}\end{equation}
 with $\mathsf{sinr}_{q}(k)$ denoting the Signal-to-Interference
plus Noise Ratio (SINR) of link $q$ on the $k$-th subchannel: \begin{equation}
\mathsf{sinr}_{q}(k)\triangleq\frac{\left\vert H_{qq}(k)\right\vert ^{2}p_{q}(k)}{\sigma_{_{q}}^{2}(k)+\sum_{\, r\neq q}\left\vert H_{qr}(k)\right\vert ^{2}p_{r}(k)},\label{SINR_q}\end{equation}
 where $\left\vert H_{qr}(k)\right\vert ^{2}$ is the power gain of
the channel between destination $q$ and source $r$; $\sigma_{_{q}}^{2}(k)$
is the variance of Gaussian zero mean noise on subchannel $k$ of
receiver $q$; and $\mathbf{p}_{q}\triangleq\left(p_{q}(k)\right)_{k=1}^{N}\mathbf{\ }$is
the power allocation strategy of user $q$ across the $N$ subchannel,
whereas $\mathbf{p}_{-q}\triangleq\left(\mathbf{p}_{r}\right)_{r\neq q}$
contains the strategies of all the other users.

\subsection{Game theoretic formulation}

\label{Sec:Problem-Formulation}We formulate the system design within
the framework of game theory \cite{Osborne,Aubin-book}, using as
desirability criterion the concept of GNE, see for example \cite{Nash-paper,Rosen}.
Specifically, we consider a strategic non-cooperative game, in which
the players are the links and the payoff functions are the transmit
powers of the users: Each player competes 
 against the others by choosing the power allocation (i.e., its strategy)
that minimizes its own transmit power, given a constraint on the minimum
achievable information rate on the link. A GNE of the game is reached
when each user, given the strategy profile of the others, does not
get any power decrease by unilaterally changing its own strategy,
still keeping the rate constraint satisfied. Stated in mathematical
terms, the game has the following structure:
\begin{equation}
{\mathscr{G}}=\left\{ \Omega,\left\{ {\mathscr{P}}_{q}(\mathbf{p}_{-q})\right\} _{q\in\Omega},\{{P}_{q}\}_{q\in\Omega}\right\} ,\label{Game G}\end{equation}
 where $\ \Omega\triangleq\left\{ 1,2,\ldots,Q\right\} $ denotes
the set of the active links, ${\mathscr{P}}_{q}(\mathbf{p}_{-q})\subseteq\mathbb{R}_{+}^{N}$
is the set of admissible power allocation strategies $\mathbf{p}_{q}\in{\mathscr{P}}_{q}(\mathbf{p}_{-q})$
of user $q$ over the subchannels $\mathcal{N}\triangleq\{1,\ldots,N\}$,
defined as\begin{equation}
\hspace{-0.2cm}{\mathscr{P}}_{q}(\mathbf{p}_{-q})\triangleq\!\left\{ \mathbf{x}_{q}\!\!\in\!\!\mathcal{\ \mathbb{R}}_{+}^{N}:\quad R_{q}(\mathbf{x}_{q},\mathbf{p}_{-q})\geq R_{q}^{\star}\right\} .\label{admissible strategy set_user_q}\end{equation}
 with $R_{q}(\mathbf{p}_{q},\mathbf{p}_{-q})$ given in (\ref{Rate}),
and $R_{q}^{\star}$ denotes the minimum transmission rate required
by each user, which we assume positive without loss of generality.
In the sequel we will make reference to the vector $\mathbf{R}^{\star}\triangleq(R_{q}^{\star})_{q=1}^{Q}$
as to the \emph{rate profile}. The payoff function of the $q$-th
player is its own transmit power $P_{q}$, given in (\ref{Tx-power}).
Observe that, because of the rate constraints, the set of feasible
strategies ${\mathscr{P}}_{q}(\mathbf{p}_{-q})$ of each player $q$
depends on the power allocations $\mathbf{p}_{-q}$ of all the other
users.

The optimal strategy for the $q$-th player, given the power allocation
of the others, is then the solution to the following minimization
problem \begin{equation}
\begin{array}{ll}
\operatorname*{minimize}\limits _{\mathbf{p}_{q}} & \quad{\displaystyle \sum\limits _{k=1}^{N}}p_{q}(k)\\
\operatorname*{subject}\text{ }\operatorname*{to} & \quad\mathbf{p}_{q}\in{\mathscr{P}}_{q}(\mathbf{p}_{-q})\end{array},\label{Power Game}\end{equation}
 where ${\mathscr{P}}_{q}(\mathbf{p}_{-q})$\ is given in (\ref{admissible strategy set_user_q}).
Note that, for each $q$, the minimum in (\ref{Power Game}) is taken
over $\mathbf{p}_{q},$ for a \textit{fixed} but arbitrary $\mathbf{p}_{-q}.$
Interestingly, given $\mathbf{p}_{-q},$ the solution of (\ref{Power Game})
can be obtained in {}``closed\textquotedblright\ form via the solution
of a singly-constrained optimization problem; see \cite{Palomar-Fonollosa05}
for an algorithm to implement this solution in practice.

\begin{lemma} \label{Lemma_WF_solution}For any fixed and nonnegative
$\mathbf{p}_{-q},$ the optimal solution $\mathbf{p}_{q}^{\star}=\{p_{q}^{\star}(k)\}_{k=1}^{N}$
of the optimization problem (\ref{Power Game}) exists and is unique.
Furthermore, \begin{equation}
\begin{array}{c}
\mathbf{p}_{q}^{\star}=\mathsf{WF}_{q}\left(\mathbf{p}_{1},\ldots,\mathbf{p}_{q-1},\mathbf{p}_{q+1},\ldots,\mathbf{p}_{Q}\right)=\mathsf{WF}_{q}(\mathbf{p}_{-q})\end{array},\quad\label{WF_single-user}\end{equation}
 where the waterfilling operator $\mathsf{WF}_{q}\left(\mathbf{\cdot}\right)$
is defined as \begin{equation}
\left[\mathsf{WF}_{q}\left(\mathbf{p}_{-q}\right)\right]_{k}\triangleq\left(\lambda_{q}-\dfrac{\sigma_{q}^{2}(k)+{\displaystyle \sum\nolimits _{\, r\neq q}}\left\vert H_{qr}(k)\right\vert ^{2}p_{r}(k)}{\left\vert H_{qq}(k)\right\vert ^{2}}\right)^{+},\quad k\in\mathcal{N},\label{WF_operator}\end{equation}
 with $\left(x\right)^{+}\triangleq\max(0,x)$ and the water-level
$\lambda_{q}$ chosen to satisfy the rate constraint $R_{q}(\mathbf{p}_{q}^{\star},\mathbf{p}_{-q})=R_{q}^{\star}$,
with $R_{q}(\mathbf{p}_{q},\mathbf{p}_{-q})$ given in (\ref{Rate}).
\end{lemma}


The solutions of the game ${\mathscr{G}}$ in $($\ref{Game G}$)$,
if they exist, are the Generalized Nash Equilibria, formally defined
as follows.

\begin{definition} \label{NE def} A feasible strategy profile $\mathbf{p}^{\star}=(\mathbf{p}_{q}^{\star})_{q=1}^{Q}$
is a GNE of the game ${\mathscr{G}}$ 
if \begin{equation}
\
{\displaystyle \sum\limits _{k=1}^{N}}p_{q}^{\star}(k)\leq{\displaystyle \sum\limits _{k=1}^{N}}p_{q}(k),\ \text{\ \ }\forall\mathbf{p}_{q}\in{\mathscr{P}}_{q}(\mathbf{p}_{-q}^{\star}),\text{ }\forall q\in\Omega.\label{pure-NE}\end{equation}

\end{definition}

According to Lemma \ref{Lemma_WF_solution}, all the Generalized Nash
Equilibria of the game must satisfy the condition expressed by the
following Corollary.

\begin{corollary} \label{Corollary_SWF_system}A feasible strategy
profile $\mathbf{p}^{\star}=(\mathbf{p}_{q}^{\star})_{q=1}^{Q}$ is
a GNE of the game ${\mathscr{G}}$ 
if and only if it satisfies the following system of nonlinear equations\begin{equation}
\begin{array}{c}
\mathbf{p}_{q}^{\star}=\mathsf{WF}_{q}\left(\mathbf{p}_{1}^{\star},\ldots,\mathbf{p}_{q-1}^{\star},\mathbf{p}_{q+1}^{\star},\ldots,\mathbf{p}_{Q}^{\star}\right)\end{array},\quad\forall q\in\Omega,\label{SWF_system}\end{equation}
 with $\mathsf{WF}_{q}\left(\mathbf{\cdot}\right)$ defined in (\ref{WF_operator}).
\end{corollary}

Given the nonlinear system of equations (\ref{SWF_system}), the fundamental
questions we want an answer to are: i) \emph{Does a solution exist,
for any given users' rate profile}? ii) \emph{If a solution exists,
is it unique}? iii) \emph{How can such a solution be reached in a
totally distributed way}?

An answer to the above questions is given in the forthcoming sections.

\section{Existence and Uniqueness of a Generalized Nash Equilibrium}

\label{Sec:Existence_Uniqueness} In this section we first provide
sufficient conditions for the existence of a nonempty and bounded
solution set of the Nash equilibrium problem (\ref{Game G}). Then,
we focus on the uniqueness of the equilibrium.

\subsection{Existence of a generalized Nash equilibrium}

Given the rate profile ${\mathbf{R}}^{\star}=(R_{q}^{\star})_{q=1}^{Q}$,
define, for each $k\in\mathcal{N}$, the matrix $\mathbf{Z}_{k}({\mathbf{R}}^{\star})\in\,\mathbb{R}^{Q\times Q}\,$
as \begin{equation}
\mathbf{Z}_{k}({\mathbf{R}}^{\star})\,\triangleq\,\left[\begin{array}{cccc}
\left\vert H_{11}(k)\right\vert ^{2}\, & -(e^{R_{1}^{\star}}-1)\left\vert H_{12}(k)\right\vert ^{2} & \cdots & -(e^{R_{1}^{\star}}-1)\,\left\vert H_{1Q}(k)\right\vert ^{2}\\[7pt]
-(e^{R_{2}^{\star}}-1)\,\left\vert H_{21}(k)\right\vert ^{2}\, & \left\vert H_{22}(k)\right\vert ^{2}\, & \cdots & -(e^{R_{2}^{\star}}-1)\,\left\vert H_{2Q}(k)\right\vert ^{2}\,\\[7pt]
\vdots & \vdots & \ddots & \vdots\\[7pt]
-(e^{R_{Q}^{\star}}-1)\,\left\vert H_{Q1}(k)\right\vert ^{2}\, & -(e^{R_{Q}^{\star}}-1)\,\left\vert H_{Q2}(k)\right\vert ^{2}\, & \cdots & \left\vert H_{QQ}(k)\right\vert ^{2}\,\end{array}\right]\,.\label{Z_kL}\end{equation}
We also need the definition of P-matrix, as given next.
\begin{definition}
A matrix $\mathbf{A}\in\mathbb{R}^{N\times N}$ is called Z-matrix if its off-diagonal entries are all non-
positive. A matrix $\mathbf{A}\in\mathbb{R}^{N\times N}$ is called P-matrix
if every principal minor of $\mathbf{A}$ is positive.
\end{definition}
Many equivalent characterizations for a P-matrix can be given. The
interested reader is referred to \cite{BPlemmons79,CPStone92} for
more details. Here we note only that any positive definite matrix
is a P-matrix, but the reverse does not hold. 

Sufficient conditions for the nonemptiness of a bounded solution set
for the game ${\mathscr{G}}$ 
are given in the following theorem.

\begin{theorem} \label{th:existence(main_body)} The game ${\mathscr{G}}$
with rate profile ${\mathbf{R}}^{\star}=(R_{q}^{\star})_{q=1}^{Q}>\mathbf{0}$
admits a nonempty and bounded solution set if $\mathbf{Z}_{k}({\mathbf{R}}^{\star})$
is a P-matrix, %
for all $k\in\mathcal{N}$, with $\mathbf{Z}_{k}({\mathbf{R}}^{\star})$
defined in (\ref{Z_kL}). Moreover, any GNE $\mathbf{p}^{\star}=(\mathbf{p}_{q}^{\ast})_{q=1}^{Q}$
is such that\begin{equation}
\left(\begin{array}{c}
p_{1}^{\ast}(k)\\[5pt]
\vdots\\[5pt]
p_{Q}^{\ast}(k)\end{array}\right)\,\leq\,\left(\begin{array}{c}
\overline{p}_{1}(k)\\[5pt]
\vdots\\[5pt]
\overline{p}_{Q}(k)\end{array}\right)\,\triangleq\,(\,\mathbf{Z}_{k}({\mathbf{R}}^{\star})\,)^{-1}\left(\begin{array}{c}
\sigma_{1}^{2}(k)\,(\, e^{R_{1}^{\star}}-1\,)\\[5pt]
\vdots\\[5pt]
\sigma_{Q}^{2}(k)\,(\, e^{R_{Q}^{\star}}-1\,)\end{array}\right),\hspace{1.0538pc}k\,\in\mathcal{N}.\label{Upper_bound_GNE}\end{equation}

\end{theorem}

\begin{proof} See Appendix \ref{proof_th:existence(main_body)}.
\end{proof}

A more general (but less easy to check) result on the existence of
a bounded solution set for the game ${\mathscr{G}}$ 
is given by Theorem \ref{th:main existence} in Appendix \ref{proof_th:existence(main_body)}.

We now provide alternative sufficient conditions for Theorem \ref{th:existence(main_body)}
in terms of a single matrix. To this end, we first introduce the following
matrix\begin{equation}
{\mathbf{Z}}^{\max}({\mathbf{R}}^{\star})\triangleq\,\left[\begin{array}{cccc}
1 & -(e^{R_{1}^{\star}}-1)\beta_{12}^{\max} & \cdots & -(e^{R_{1}^{\star}}-1)\,\beta_{1Q}^{\max}\\[0.2in]
-(e^{R_{2}^{\star}}-1)\,\beta_{21}^{\max} & 1 & \cdots & -(e^{R_{2}^{\star}}-1)\beta_{2Q}^{\max}\\[0.2in]
\vdots & \vdots & \ddots & \vdots\\[7pt]
-(e^{R_{Q}^{\star}}-1)\,\beta_{Q1}^{\max} & -(e^{R_{Q}^{\star}}-1)\,\beta_{Q2}^{\max} & \cdots & 1\end{array}\right]\,,\label{Z_max}\end{equation}
 where \begin{equation}
\beta_{qr}^{\max}\,\triangleq\max\limits _{k\in\mathcal{N}}\,\dfrac{\left\vert H_{qr}(k)\right\vert ^{2}}{\left\vert H_{rr}(k)\right\vert ^{2}},\hspace{1.0152pc}\forall r\neq q{\,,\quad q\in\Omega.}\label{beta_max}\end{equation}
 We also denote by $e^{{\mathbf{R}}^{\star}}-\mbox{\boldmath{$1$}}$
the $Q$-vector with $q$-th component $e^{R_{q}^{\star}}-1,$ for
$q=1,\ldots,Q$. Then, we have the following corollary.

\begin{corollary} \label{Corollary:SF_Existence_Z_maxL} 

If ${\mathbf{Z}}^{\max}({\mathbf{R}}^{\star})$ in (\ref{Z_max})
is a P-matrix, then all the matrices $\{\mathbf{Z}_{k}({\mathbf{R}}^{\star})\}$
defined in (\ref{Z_kL}) are P-matrices. Moreover, any GNE $\mathbf{p}^{\star}=(\mathbf{p}_{1}^{\star})_{q=1}^{Q}$
of the game ${\mathscr{G}}$ 
satisfies\begin{equation}
p_{q}^{\ast}(k)\leq\overline{p}_{q}(k)=\left(\frac{\max\limits _{r\in\Omega}\sigma_{r}^{2}(k)}{\left\vert H_{qq}(k)\right\vert ^{2}}\right)\left[\left({\mathbf{Z}}^{\max}({\mathbf{R}}^{\star})\right)^{-1}\left(e^{{\mathbf{R}}^{\star}}-\mbox{\boldmath{$1$}}\right)\right]_{q},\quad\forall q\in\Omega,\quad\forall k\,\in\mathcal{N}.\label{eq:FFbounds}\end{equation}

\end{corollary}

\begin{proof} See Appendix \ref{proof_Corollary:SF_Existence_Z_maxL}.
\end{proof}

To give additional insight into the physical interpretation of the
existence conditions of a GNE, we make explicit the dependence of
each channel (power) gain $|H_{qr}(k)|^{2}$ on its own source-destination
distance $d_{qr}$ by introducing the normalized channel gain $|\overline{H}_{qr}(k)|^{2}=|H_{qr}(k)|^{2}d_{qr}^{\gamma},$
where $\gamma$ is the path loss exponent. We have the following corollary.

\begin{corollary} \label{Corollary:SF_Existence_for_ZkL}Sufficient
conditions for the matrices $\{\mathbf{Z}_{k}({\mathbf{R}}^{\star})\}$
defined in (\ref{Z_kL}) to be P-matrices are:\begin{equation}
{\displaystyle \sum\limits _{r\neq q}}\dfrac{|\overline{H}_{qr}(k)|^{2}}{|\overline{H}_{qq}(k)|^{2}}\dfrac{d_{qq}^{\,\gamma}}{d_{qr}^{\,\gamma}}<\frac{1}{\, e^{R_{q}^{\star}}-1\,},\quad\quad\forall r\in\Omega,\quad\forall k\,\in\mathcal{N}.\label{Corollary_SF_for_Z_kL}\end{equation}

\end{corollary}

\begin{proof} The proof comes directly from the sufficiency of the
diagonally dominance property \cite[Definition 2.2.19]{CPStone92} for the matrices $\mathbf{Z}_{k}({\mathbf{R}}^{\star})$
in (\ref{Z_kL}) to be P-matrices \cite[Theorem 6.2.3]{BPlemmons79} 
\end{proof}

\begin{remark}\rm A physical interpretation of the conditions in Theorem
\ref{th:existence(main_body)} (or Corollary \ref{Corollary:SF_Existence_for_ZkL})
is the following. Given the set of channels and
the rate constraints, a GNE of $\mathscr{G}$ is guaranteed to exist
if multiuser interference is {}``sufficiently small\textquotedblright$\,$
(e.g., the links are sufficiently far apart). In fact, from (\ref{Corollary_SF_for_Z_kL}),
which quantifies the concept of small interference, one infers that,
for any fixed set of (normalized) channels and rate constraints, there exists a
minimum distance beyond which an equilibrium exists, corresponding
to the maximum level of interference that may be tolerated from each
user. The amount of such a tolerable multiuser interference depends
on the rate constraints: the larger the required rate from each user,
the lower the level of interference guaranteeing the existence of
a solution. The reason why an equilibrium of the game $\mathscr{G}$
might not exist for any given set of channels and rate constraints,
is that the multiuser system we consider is interference limited,
and thus not every QoS requirement is guaranteed to be feasible. In
fact, in the game $\mathscr{G}$, each user acts to increase the transmit
power to satisfy its own rate constraint; which leads to an increase
of the interference against the users. It turns out that, increasing
the transmit power of all the users does not guarantee that an equilibrium
could exist for any given rate profile.

Observe that conditions in Theorem \ref{th:existence(main_body)}
also provide a simple admission control procedure to check if a set
of rate constraints is feasible: under these conditions indeed, one
can always find a \emph{finite} power budget for all the users such
that there exists a GNE where all the rate constraints are satisfied.
\end{remark}

\subsection{ Uniqueness of the Generalized Nash Equilibrium}

Before providing conditions for the uniqueness of the GNE of the game
$\mathscr{G}$, we introduce the following intermediate definitions.
For any given rate profile ${\mathbf{R}}^{\star}=(R_{q}^{\star})_{q=1}^{Q}>\mathbf{0},$
let $\overline{\mathbf{B}}({\mathbf{R}}^{\star})\in\,\mathbb{R}^{Q\times Q}\,$
be defined as\begin{equation}
\left[\overline{\mathbf{B}}({\mathbf{R}}^{\star})\right]_{qr}\,\equiv\,\left\{ \begin{array}{ll}
e^{-R_{q}^{\star}}, & \quad\text{if }q=r,\\[5pt]
-e^{R_{q}^{\star}}\,\widehat{\beta}_{qr}^{\max}, & \quad\text{otherwise,}\end{array}\right.\label{Beta_bar_matrix}\end{equation}

\noindent where\begin{equation}
\widehat{\beta}_{qr}^{\max}\,\triangleq\max\limits _{k\,\in\mathcal{N}}\left(\,\dfrac{\left\vert H_{qr}(k)\right\vert ^{2}}{\left\vert H_{rr}(k)\right\vert ^{2}}\dfrac{\sigma_{r}^{2}(k)+\sum_{\, r^{\,\prime}\neq r}\left\vert H_{rr^{\,\prime}}(k)\right\vert ^{2}\overline{p}_{r^{\,\prime}}(k)}{\sigma_{q}^{2}(k)}\right),\label{eq:def:beta_max_hat}\end{equation}

\noindent with $\overline{p}_{r^{\,\prime}}(k)$ defined in (\ref{Upper_bound_GNE}).
We also introduce $\chi$ and $\rho,$ defined respectively as\begin{equation}
\chi\triangleq1-\max_{q\in\Omega}\left[\left(\, e^{R_{q}^{\star}}-1\right){\displaystyle \sum\limits _{r\neq q}}\beta_{qr}^{\max}\,\right],\label{eq:def:_xi}\end{equation}

\noindent with $\beta_{qr}^{\max}$ given in (\ref{beta_max}), and\begin{equation}
\rho\triangleq\frac{e^{R_{\max}^{\star}}-1}{e^{R_{\min}^{\star}}-1},\label{eq:def_rho}\end{equation}

\noindent with\begin{equation}
R_{\max}^{\star}\triangleq\max\limits _{q\in\Omega}R_{q}^{\star},\text{\quad and\quad}R_{\min}^{\star}\triangleq\min\limits _{q\in\Omega}R_{q}^{\star}.\end{equation}

Sufficient conditions for the uniqueness of the GNE of the game $\mathscr{G}$
are given in the following theorem.

\begin{theorem} \label{th:uniqueness_(main body)} Given the game
${\mathscr{G}}$ 
with a rate profile ${\mathbf{R}}^{\star}=(R_{q}^{\star})_{q=1}^{Q}>\mathbf{0},$
assume that the conditions of Theorem \ref{th:existence(main_body)}
are satisfied. 
If, in addition, $\overline{\mathbf{B}}({\mathbf{R}}^{\star})$ in
(\ref{Beta_bar_matrix}) is a P-matrix, then the game ${\mathscr{G}}$
admits a unique GNE. \end{theorem}

\begin{proof} See Appendix \ref{proof_th:uniqueness_(main body)}.
\end{proof}

More stringent but more intuitive conditions for the uniqueness of
the GNE are given in the following corollary.

\begin{corollary} \label{Corollary:SF_Uniqueness}Given the game
${\mathscr{G}}$ with rate profile ${\mathbf{R}}^{\star}=(R_{q}^{\star})_{q=1}^{Q}>\mathbf{0},$
assume that \begin{equation}
0<\chi<1,\end{equation}
 so that a GNE for the game ${\mathscr{G}}$ is guaranteed to exist,
with $\chi$ defined in (\ref{eq:def:_xi}). Then, the GNE is unique
if the following conditions hold true\begin{equation}
{\sum_{r\neq q}}\,\beta_{qr}^{\max}\,\,\left\{ \,\left({\max_{k\,\in\mathcal{N}}\frac{\sigma_{r}^{2}(k)}{\sigma_{q}^{2}(k)}}\right)\,+\left[{\max_{r^{\,\prime}\in\Omega}}\,\left({\max_{k\,\in\mathcal{N}}}\,{\frac{\sigma_{r^{\,\prime}}^{2}(k)}{\sigma_{q}^{2}(k)}}\right)\right]\,\left({\frac{\rho}{\chi}}-1\,\right)\,\right\} <\frac{1}{e^{2R_{q}^{\star}}},\quad\forall q\in\Omega,\label{SF_2_Uniqueness}\end{equation}

\noindent with $\rho$ defined in (\ref{eq:def_rho}).

In particular, when $\sigma_{r}^{2}(k)=\sigma_{q}^{2}(n),$ $\forall r,q\in\Omega$
and $\forall k,n\in\mathcal{N},$ conditions (\ref{SF_2_Uniqueness})
become\begin{equation}
{\displaystyle \sum\limits _{r\neq q}}\max\limits _{k\,\in\mathcal{N}}\left\{ \dfrac{|\overline{H}_{qr}(k)|^{2}}{|\bar{H}_{rr}(k)|^{2}}\right\} \dfrac{d_{rr}^{\,\gamma}}{d_{qr}^{\,\gamma}}\,\,<\frac{\,\gamma}{e^{R_{q}^{\star}}-1},\quad\forall q\in\Omega,\label{eq:SF_uniq_equal_sigmas}\end{equation}

\noindent with \begin{equation}
\gamma\triangleq\frac{{\max\limits _{q\in\Omega}}\left\{ e^{-R_{q}^{\star}}-e^{-2R_{q}^{\star}}\right\} }{\dfrac{e^{R_{\max}^{\star}}-1}{e^{R_{\min}^{\star}}-1}+{\max\limits _{q\in\Omega}}\left\{ e^{-R_{q}^{\star}}-e^{-2R_{q}^{\star}}\right\} }<1.\end{equation}

\end{corollary}

\begin{proof} See Appendix \ref{proof_Corollary:SF_Uniqueness}.
\end{proof}

\subsection{On the conditions for existence and uniqueness of the GNE}

It is natural to ask whether  the sufficient conditions as given by
Theorem \ref{th:existence(main_body)} (or the more general ones given
by Theorem \ref{th:main existence} in Appendix A) are tight. In the next proposition, 
we show that these conditions become
indeed necessary in the special case of $N=1$ subchannel.  


\begin{proposition} \label{pr:1-tone problem} Given the rate profile
$\mathbf{R}^{\star}=(R_{q}^{\star})_{q=1}^{Q},$ the following 
statements are equivalent for the game $\mathscr{G}$ when $N=1$:%
\footnote{In the case of $N=1$, the power allocation $p_{q}(k)=p_{q}$ of each
user, the channel gains $\left\vert H_{rq}(k)\right\vert ^{2}=\left\vert H_{rq}\right\vert ^{2}$
and the noise variances $\sigma_{q}^{2}(k)=\sigma_{q}^{2}$ are independent
on index $k$. Matrix $\mathbf{Z}_{k}({\mathbf{R}}^{\star})=\mathbf{Z}({\mathbf{R}}^{\star})$
is defined as in (\ref{Z_kL}), where each $\left\vert H_{rq}(k)\right\vert ^{2}$
is replaced by  $\left\vert H_{rq}\right\vert ^{2}.$ %
}

\begin{description}
\item [{\rm}] (a) The problem (\ref{Power Game}) has a solution for some
(all) $(\sigma_{q}^{2})_{q=1}^{Q}>0$. 
\item [{\rm}] (b) The matrix $\mathbf{Z}({\mathbf{R}}^{\star})$ is a
P-matrix. 
\end{description}
If any one of the above two statements holds, then the game $\mathscr{G}$
has a unique solution that is the unique solution to the system of
linear equations: \begin{equation}
\left\vert H_{qq}\right\vert ^{2}\, p_{q}-(\, e^{R_{q}^{\star}}-1\,)\,{\displaystyle {\sum_{r\neq q}}\,\left\vert H_{qr}\right\vert ^{2}\, p_{r}\,=\,\sigma_{q}^{2}\,(\, e^{R_{q}^{\star}}-1\,)\qquad\forall q\in\Omega.}\label{eq:1-tone}\end{equation}
 \end{proposition}

\begin{proof} See Appendix \ref{proof_Proposition_G_one_tone}. \end{proof}

\begin{remark} \rm Proposition \ref{pr:1-tone problem} also shows that it is, in general, very hard to obtain improved sufficient condition for the existence and boundedness of solutions to the problem with $N > 1$, as any such condition must be implied by condition (b) above for the the 1-subchannel case, which, as shown by the proposition, is necessary for the said existence, and also for the uniqueness as it turns out.
\end{remark}
%

\begin{figure}[h]
\vspace{-0.5cm}
\par
\begin{center}
\includegraphics[trim=0.000000in 0.000000in 0.000000in
-0.212435in, width=7cm,height=7cm]{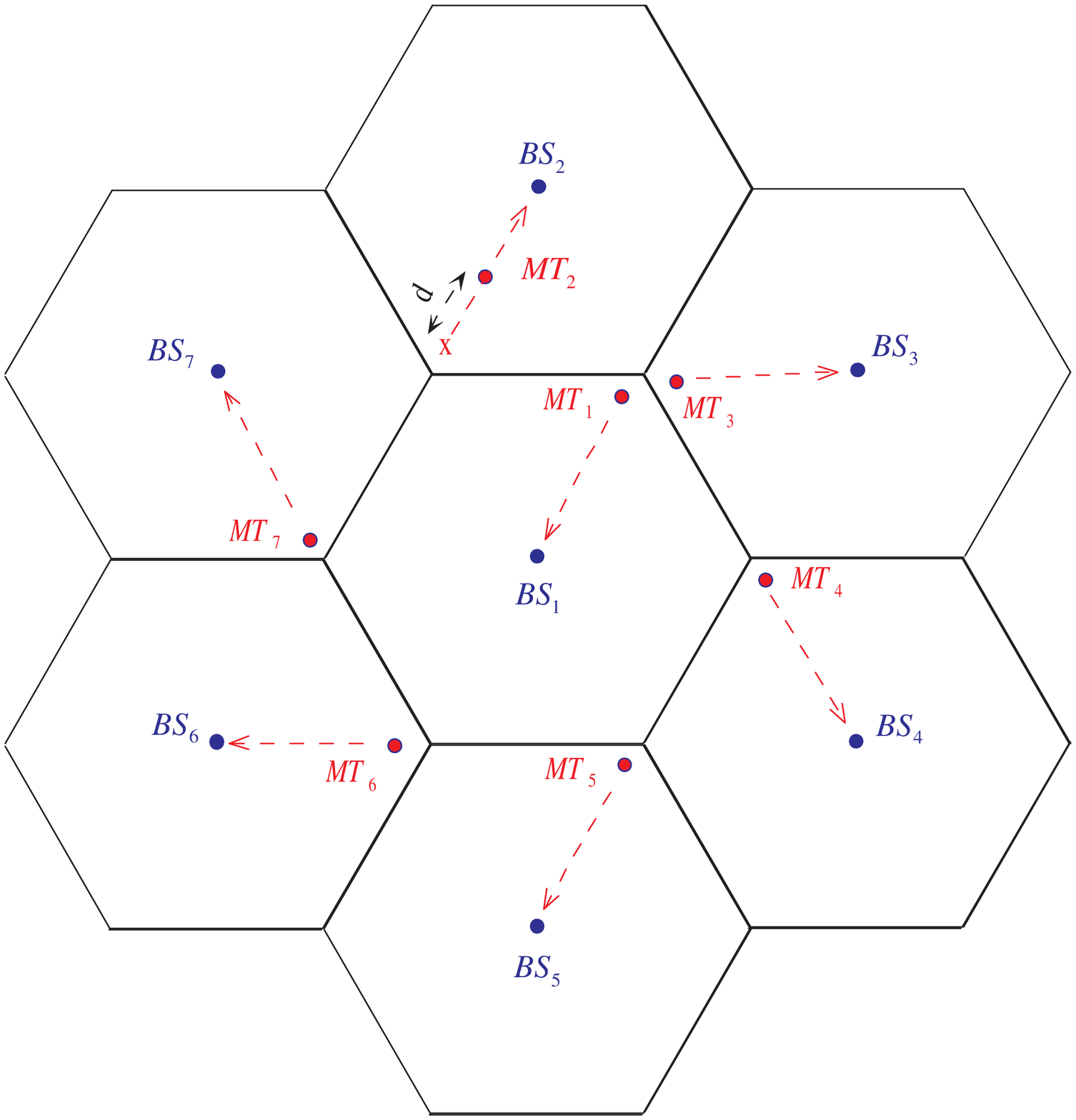}

(a)
\par
\includegraphics[trim=0.000000in 0.000000in 0.000000in
-0.212435in, width=11cm,height=7.7cm]{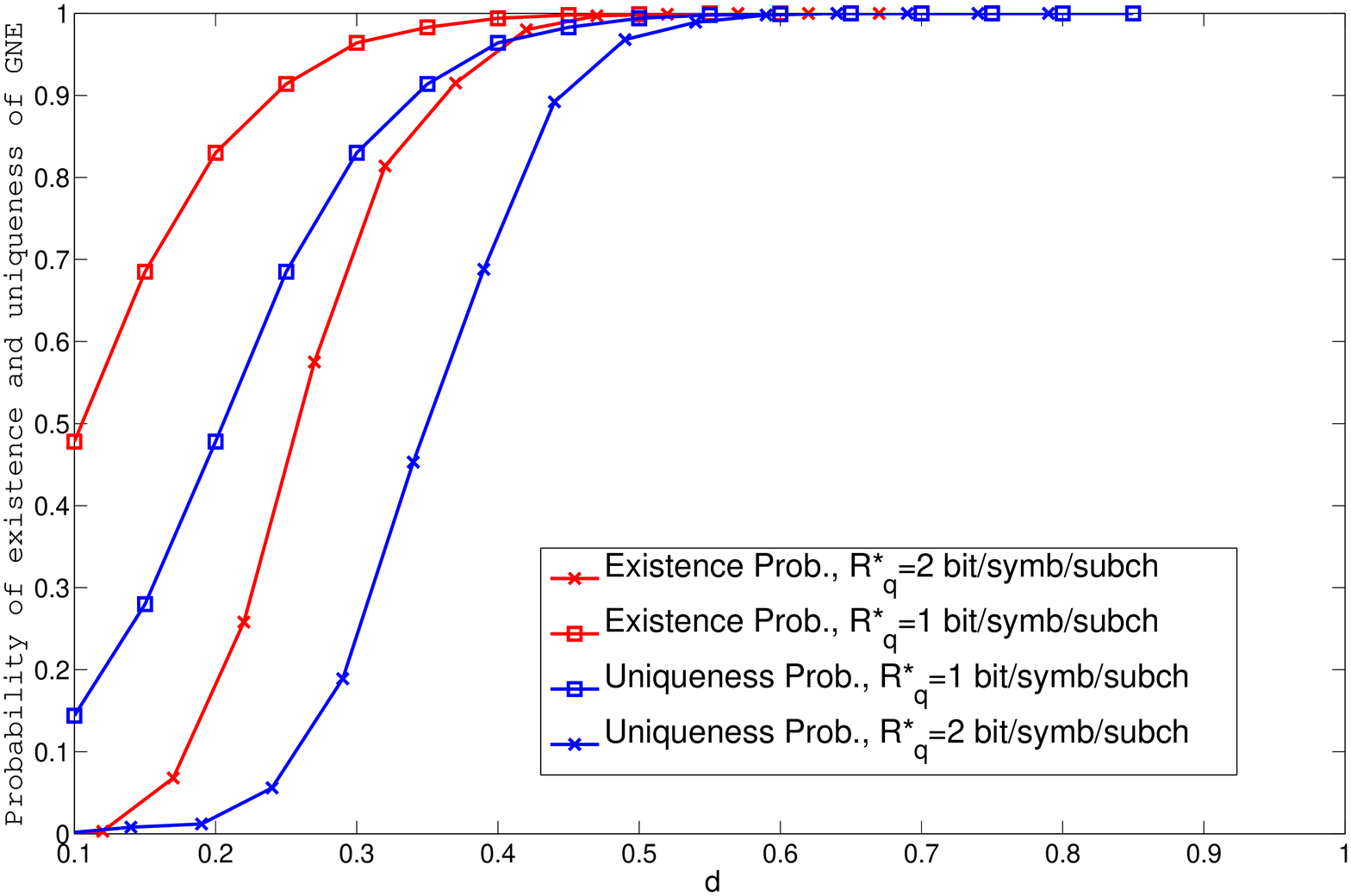}

\vspace{-0.4cm}(b)
\end{center}
\par
\vspace{-0.8cm}\caption{{\protect{\small Probability of existence (red line curves)
and uniqueness (blue line curves) of the GNE
versus $d$ {[}subplot (b)] for a $7$-cell (downlink) cellular system
{[}subplot (a)] and rate profiles $R_{q}^{\star}=1$ bit/symb/subchannel
(square markers) and $R_{q}^{\star}=2$ bit/symb/subchannel (cross
markers), $\forall q\in\Omega$.}}}\label{Fig:check_cond}
\end{figure}

\begin{remark} \rm Observe that, when $N=1$, the game $\mathscr{G}$
leads to classical SINR based \emph{scalar }power control problems
in flat-fading CDMA (or TDMA/FDMA) systems, where the goal of each
user is to reach a prescribed SINR (see (\ref{SINR_q})) with the
minimum transmit power $P_{q}$ \cite{Bambos}. In this case, given
the rate profile $\mathbf{R}^{\star}=(R_{q}^{\star})_{q=1}^{Q}$ and
$N=1,$ the SINR target profile $\boldsymbol{\mathsf{sinr}}^{\star}\triangleq(\mathsf{sinr}_{q}^{\star})_{q=1}^{Q},$
as required in classical power control problems \cite{Bambos}, can
be equivalently written in terms of $\mathbf{R}^{\star}$as\begin{equation}
\mathsf{sinr}_{q}^{\star}=e^{-R_{q}^{\star}}-1,\quad q\in\Omega,\end{equation}
 and the Nash equilibria $\mathbf{p}^{\star}=(p_{q}^{\star})_{q=1}^{Q}$
of the game $\mathscr{G}$ become the solutions of the following system
of linear equations\begin{equation}
\,\,\mathbf{Z}({\mathbf{R}}^{\star})\mathbf{p}^{\star}=\left(\begin{array}{c}
\sigma_{1}^{2}\,\mathsf{sinr}_{1}^{\star}\\[5pt]
\vdots\\[5pt]
\sigma_{Q}^{2}\mathsf{sinr}_{Q}^{\star}\end{array}\right).\label{eq:scalar_power_control_problem}\end{equation}

Interestingly,  the necessary and sufficient condition (b) given in Proposition \ref{pr:1-tone problem}
is equivalent to that known in the literature for the existence and
uniqueness of the solution of the classical SINR based power control
problem (see, e.g., \cite{Bambos}). Moreover, observe that, in the
case of $N=1,$ the solution of the game $\mathscr{G},$ coincides
with the upper bound in (\ref{Upper_bound_GNE}).\end{remark}
\medskip
\noindent \textbf{Numerical example.} 
Since the existence and uniqueness
conditions of the GNE given so far depend on the channel power gains
$\left\{ |H_{qr}(k)|^{2}\right\} $, there is a nonzero probability
that they are not satisfied for a given channel realization drawn
from a given probability space and rate profile. To quantify the adequacy
of our conditions, we tested them over a set of channel impulse responses
generated as vectors composed of $L=6$ i.i.d. complex Gaussian random
variables with zero mean and variance equal to the square distance
between the associated transmitter-receiver links (multipath Rayleigh
fading model). Each user transmits over a set of $N=32$ subcarriers.
We consider a multicell cellular network as depicted in Figure \ref{Fig:check_cond}a),
composed of $7$ (regular) hexagonal cells, sharing the same band.
Hence, the transmissions from different cells typically interfere
with each other. For the simplicity of representation, we assume that
in each cell there is only one active link, corresponding to the transmission
from the base station (BS) to a mobile terminal (MT). 
According to this geometry, each MT receives a useful signal that
is comparable, in average sense, with the interference signal transmitted
by the BSs of two adjacent cells. The overall network can be modeled
as a $7$-users interference channel, composed of $32$ parallel subchannels.
In Figure \ref{Fig:check_cond}b), we plot the probability that existence
(red line curves) and uniqueness (blue line curves) conditions as
given in Theorem \ref{th:existence(main_body)} and Theorem \ref{th:uniqueness_(main body)},
respectively, are satisfied versus the (normalized) distance $d\in[0,1)$
[see Figure \ref{Fig:check_cond}a)], between each MT and his BS (assumed
to be equal for all the MT/BS pairs). We considered two different
rate profiles, namely $R_{q}^{\star}=1$ bit/symb/subchannel (square
markers) and $R_{q}^{\star}=2$ bit/symb/subchannel (cross markers),
$\forall q\in\Omega$. As expected, the probability of existence and
uniqueness of the GNE increases as each MT approaches his BS (i.e.,
$d\rightarrow1$), corresponding to a decrease of the intercell interference.
%
%
%
%
%

\section{Distributed Algorithms}

\label{Sec:IWFAs} The game $\mathscr{G}$ was shown to admit a GNE,
under some technical conditions, where each user attains the desired
information rate with the minimum transmit power, given the power
allocations at the equilibrium of the others. In this section, we
focus on algorithms to compute these solutions. Since we are interested
in a decentralized implementation, where no signaling among different
users is allowed, we consider totally distributed algorithms, where
each user acts independently to optimize its own power allocation
while perceiving the other users as interference. More specifically,
we propose two alternative totally distributed algorithms based on
the waterfilling solution in (\ref{WF_single-user}), and provide
a unified set of convergence conditions for both algorithms.\vspace{-0.3cm}

\subsection{Sequential iterative waterfilling algorithm}

The sequential Iterative Waterfilling Algorithm (IWFA) we propose
is an instance of the Gauss-Seidel scheme (by which, each user's power
is sequentially updated \cite{Bertsekas Book-Parallel-Comp}) based
on the mapping (\ref{WF_single-user}): Each player, sequentially
and according to a fixed updating order, solves problem (\ref{Power Game}),
performing the single-user waterfilling solution in (\ref{WF_single-user}).
The sequential IWFA is described in Algorithm 1.


\begin{algo}{Sequential Iterative Waterfilling Algorithm}SSet $\mathbf{p}_{q}^{(0)}=$
any nonnegative vector; \newline\texttt{for} $n=0:\mathrm{Number\hspace{0.005cm}\_\hspace{0.05cm}of\hspace{0.005cm}\_\ iterations,}$
\newline \begin{equation}
\,\,\,\,\mathbf{p}_{q}^{(n+1)}=\left\{ \begin{array}{ll}
\mathsf{WF}_{q}\left(\mathbf{p}_{-q}^{(n)}\right), & \text{if }(n+1)\,\text{mod}\, Q=q,\\
\mathbf{p}_{q}^{(n)}, & \text{otherwise},\end{array}\right.\hspace{1cm}\forall q\in\Omega;\end{equation}
 \label{IWFA_op} \newline \texttt{end}\label{IWFA_Algo} \end{algo}

\bigskip{}

The convergence of the algorithm is guaranteed under the following
sufficient conditions.

\begin{theorem} \label{th:Convergence_IWFA} Assuming $\mathrm{Number\hspace{0.005cm}\_\hspace{0.05cm}of\hspace{0.005cm}\_\ iterations}=\infty$,
the sequential IWFA, described in Algorithm 1, converges linearly
\footnote{A sequence $\{x_{n}\}$ is said to converge linearly to $x^{\star}$  if there is a constant $0<c<1$ such that  $|| x_{n+1}-x^{\star}|| \leq c || x_{n}-x^{\star}||$ for all $n\geq \bar{n}$ and some $\bar{n}\in \mathbb{N}$.} to the unique GNE of the game ${\mathscr{G}}$, 
if the conditions of Theorem \ref{th:uniqueness_(main body)} are
satisfied. \end{theorem}

\begin{proof} See Appendix \ref{proof_th:Convergence_IWFA-SIWFA}.
\end{proof}

\medskip{}

\begin{remark}\rm Observe that the convergence of the algorithm is guaranteed
under the same conditions obtained for the uniqueness of the solution
of the game. As expected, the convergence is ensured if the level
of interference in the network is not too high. \end{remark}

\begin{remark} \rm The main features of the proposed algorithm are its
low-complexity and distributed nature. In fact, despite the coupling
among the users' admissible strategies due to the rate constraints,
the algorithm can be implemented in a totally distributed way, since
each user, to compute the waterfilling solution (\ref{WF_single-user}),
only needs to locally measure the interference-plus-noise power over
the $N$ subchannels [see (\ref{SINR_q})] and waterfill over this
level. \end{remark}


\begin{remark} \rm Despite its appealing properties, the sequential IWFA
described in Algorithm $1$ may suffer from slow convergence if the
number of users in the network is large, as we will also show numerically
in Section \ref{Sec:SIWFA}. This drawback is due to the sequential
schedule in the users' updates, wherein each user, to choose its own
strategy, is forced to wait for all the other users scheduled before
it. It turns out that the sequential schedule, as in Algorithm $1$,
does not really gain from the distributed nature of the multiuser
system, where each user, in principle, is able to change its own strategy,
irrespective of the update times of the other users. Moreover, to
be performed, the sequential update requires a centralized synchronization
mechanism that determines the order and the update times of the users.
We address more precisely this issue in the next section. \end{remark}

\subsection{Simultaneous iterative waterfilling algorithm}

\label{Sec:SIWFA} To overcome the drawback of the possible slow speed
of convergence, we consider in this section the \emph{simultaneous}
version of the IWFA, called simultaneous Iterative-waterfilling Algorithm.
The algorithm is an instance of the Jacobi scheme \cite{Bertsekas Book-Parallel-Comp}:
At each iteration, the users update their own PSD \emph{simultaneously},
performing the waterfilling solution (\ref{WF_single-user}), given
the interference generated by the other users in the \emph{previous}
iteration. The simultaneous IWFA is described in Algorithm $2$.

\bigskip{}

\begin{algo}{\hspace{0.6cm} Simultaneous Iterative Waterfilling
Algorithm } SSet $\mathbf{p}_{q}^{(0)}=$ any nonnegative vector,\,
$\forall q\in\Omega;$\vspace{0.1cm}
 \newline \texttt{for} $n=0:\mathrm{Number\hspace{0.005cm}\_\hspace{0.05cm}of\hspace{0.005cm}\_\ iterations}$
\begin{equation}
\,\,\,\,\mathbf{p}_{q}^{(n+1)}={\mathsf{{WF}}}_{q}\left(\mathbf{p}_{1}^{(n)},\ldots,\mathbf{p}_{q-1}^{(n)},\mathbf{p}_{q+1}^{(n)},\ldots,\mathbf{p}_{Q}^{(n)}\right),\hspace{1cm}\forall q\in\Omega,\label{SIWFA_op}\end{equation}
 \texttt{end} \end{algo}

\bigskip{}

Interestingly, (sufficient) conditions for the convergence of the
simultaneous IWFA are the same as those required by the sequential
IWFA, as given in the following.

\begin{theorem} \label{th:Convergence_SIWFA} Assuming $\mathrm{Number\hspace{0.005cm}\_\hspace{0.05cm}of\hspace{0.005cm}\_\ iterations}=\infty$,
the simultaneous IWFA, described in Algorithm 2, converges linearly
to the unique GNE of the game ${\mathscr{G}}$, 
if the conditions of Theorem \ref{th:uniqueness_(main body)} are
satisfied. \end{theorem}

\begin{proof} See Appendix \ref{proof_th:Convergence_IWFA-SIWFA}.
\end{proof}

\begin{remark}\rm Since the simultaneous IWFA is still based on the waterfilling
solution (\ref{WF_single-user}), it keeps the most appealing features
of the sequential IWFA, namely its low-complexity and distributed
nature. In addition, thanks to the Jacobi-based update, all the users
are allowed to choose their optimal power allocation simultaneously.
Hence, the simultaneous IWFA is expected to be faster than the sequential
IWFA, especially if the number of active users in the network is large.
\end{remark}

\begin{figure}[h]
\vspace{-0.5cm}
\par
\begin{center}
\includegraphics[trim=0.000000in 0.000000in 0.000000in
-0.212435in, height=7cm]{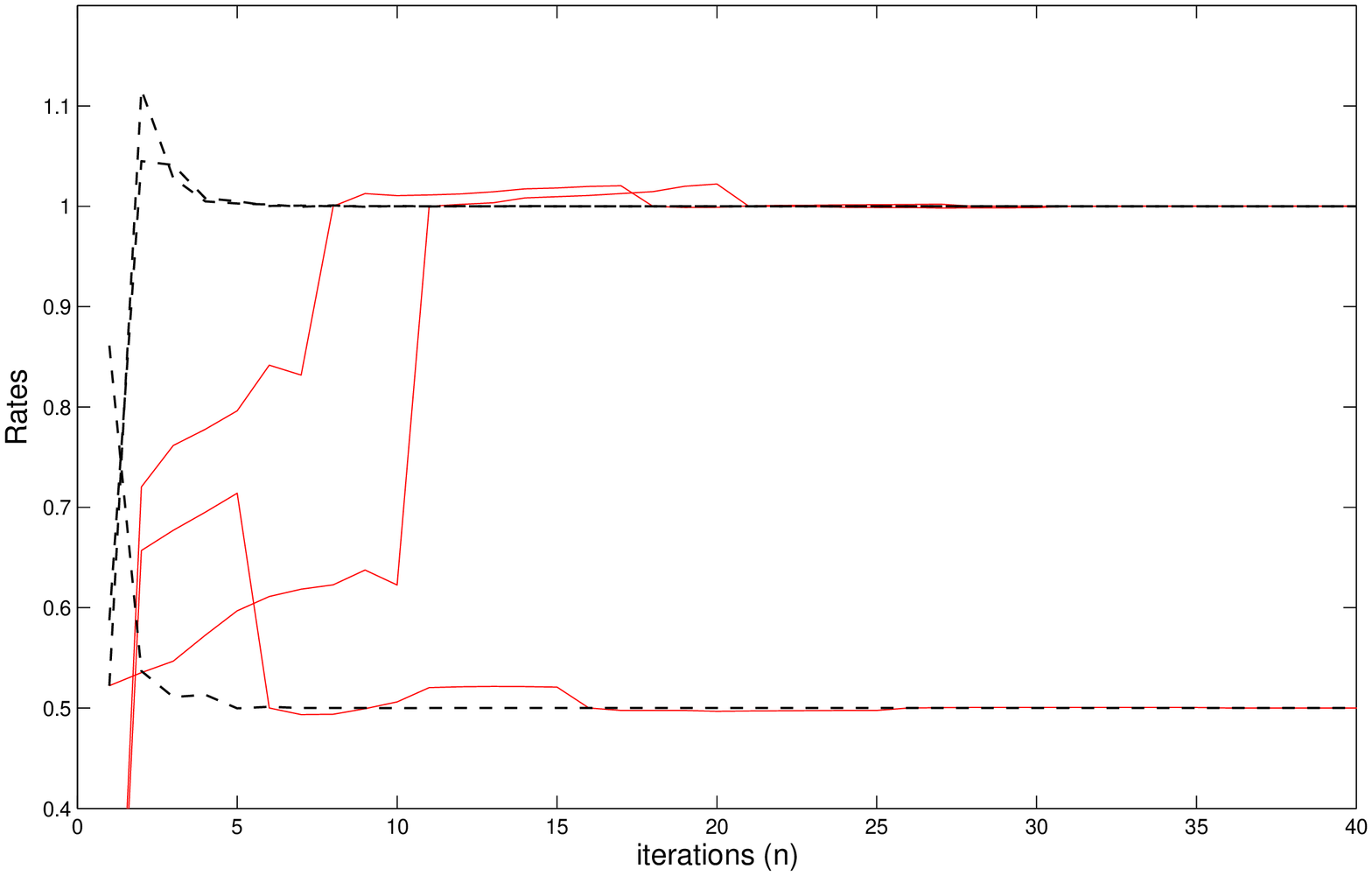}

(a)
\par
\includegraphics[trim=0.000000in 0.000000in 0.000000in
-0.212435in, height=7cm]{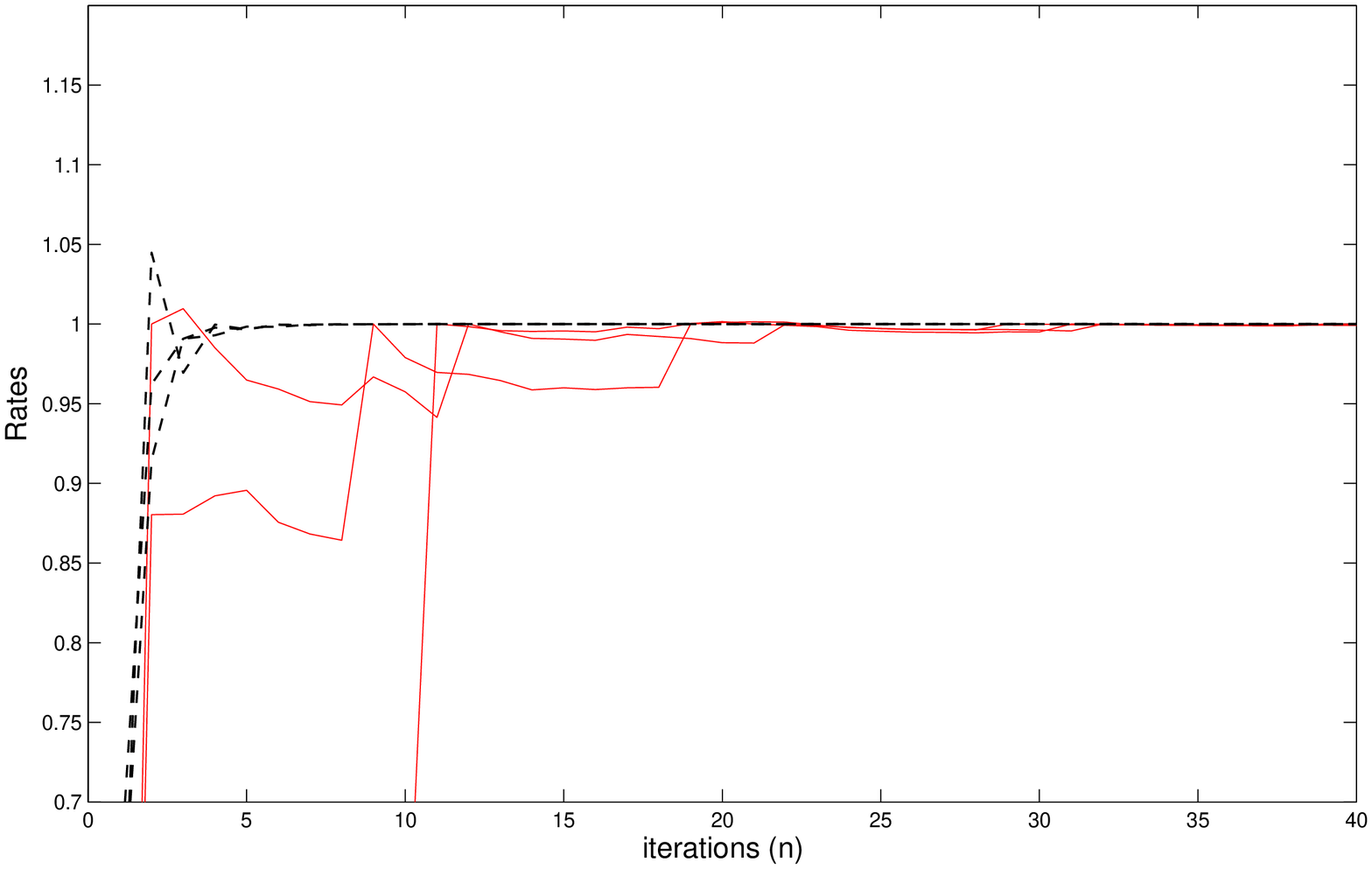}

(b)
\end{center}
\par
\vspace{-0.4cm}\vspace{-0.3cm}\caption{{{\small Rates of the users versus iterations: sequential IWFA (solid
line curves), simultaneous IWFA (dashed line curves), $Q=10,$ $d_{rq}/d_{qr}$,
$d_{rr}=d_{qq}=1,$ $\gamma=2.5$.}}}
\label{SIWFA-IWFA}%
\end{figure}

\noindent \textbf{Numerical Example.} As an example, in Figure \ref{SIWFA-IWFA},
we compare the performance of the sequential and simultaneous IWFA,
in terms of convergence speed. We consider a network composed of 10
links and we show the rate evolution of three of the links corresponding
to the sequential IWFA and simultaneous IWFA as a function of the
iteration index $n$ as defined in Algorithms 1 and 2. In Figure \ref{SIWFA-IWFA}a)
we consider a rate profile for the users with two different classes
of service; whereas in Figure \ref{SIWFA-IWFA}b) the same target
rate for all the users is required. As expected, the sequential IWFA
is slower than the simultaneous IWFA, especially if the number of
active links $Q$ is large, since each user is forced to wait for
all the other users scheduled before updating its power allocation.\vspace{-0.2cm}

\noindent %
%
%
%
%
%
%
%

\section{Conclusions}

\label{Sec:Conclusions}In this paper we have considered the distributed
power allocation in Gaussian parallel interference channels,
subject to QoS\ constraints. More specifically, we have proposed
a new game theoretic formulation of the power control problem, where
each user aims at minimizing the transmit power while guaranteeing
a prescribed information rate. We have provided sufficient conditions
for the nonemptiness and the boundedness of the solution set of the
Nash problem. These conditions suggest a simple admission control
procedure to check the feasibility of any given users' rate profile.
As expected, there exists a trade-off between the performance achievable
from each user (i.e., the achievable information rate) and the maximum level
of interference that may be tolerated in the network. Under some additional
conditions we have shown that the solution of the generalized Nash
problem is unique and we have proved the convergence of two distributed
algorithms: The sequential and the simultaneous IWFAs. Interestingly,
although the rate constraints induce a coupling among the feasible
strategies of the users, both algorithms are totally distributed,
since each user, to compute the waterfilling solution, only needs
to measure the noise-plus-interference power across the subchannels.
Our results are thus appealing in all the practical distributed multipoint-to-multipoint
systems, either wired or wireless, where centralized power control
techniques are not allowed and QoS in terms of information rate must
be guaranteed for each link.

One interesting direction that is worth of further investigations
is the generalization of the proposed algorithms to the case of asynchronous
transmission and totally asynchronous updates among the users, as
did in \cite{Scutari-Palomar-Barbarossa-IT_Journal} for the rate
maximization game. 

\section{Appendices}

\appendix
\vspace{-0.2cm}

\section{Proof of Theorem \ref{th:existence(main_body)}}

\label{proof_th:existence(main_body)}We derive Theorem \ref{th:existence(main_body)}
as a corollary to the more general Theorem \ref{th:main existence}
below. In order to prove this theorem we need several preliminary
concepts and results though, as given next.

\subsection{Noiseless game}

We rewrite first the KKT optimality conditions of the Nash problem
(\ref{Power Game}) as a Mixed nonlinear Complementarity Problem (MNCP)
\cite{FPang03,CPStone92}. Denoting by $\mu_{q,k}$ the multipliers
of the nonnegativity constraints and by $\lambda_{q}$ the multipliers
of the rate constraint, the KKT conditions of problems (\ref{Power Game})
can be written as: \begin{equation}
\begin{array}{ll}
1-\mu_{q,k}-\lambda_{q}\dfrac{\left\vert H_{qq}(k)\right\vert ^{2}}{\sigma_{_{q}}^{2}(k)+\sum_{\, r=1}^{Q}\left\vert H_{qr}(k)\right\vert ^{2}p_{r}(k)}=0, & \quad\forall k\in\mathcal{N}\text{, }\forall q\in\Omega,\\
0\,\leq\, p_{q}(k)\quad\perp\quad\mu_{q,k}\geq0, & \quad\forall k\in\mathcal{N}\text{, }\forall q\in\Omega,\\
0\,\leq\,\lambda_{q}\quad\quad\hspace{0.1cm}\perp\quad\,\,\,\,{\displaystyle \sum\limits _{k=1}^{N}}\log\left(1+\text{ }\dfrac{\left\vert H_{qq}(k)\right\vert ^{2}p_{q}(k)}{\sigma_{_{q}}^{2}(k)+\sum_{\, r\neq q}\left\vert H_{qr}(k)\right\vert ^{2}p_{r}(k)}\right)-R_{q}^{\star}\geq0, & \quad\forall q\in\Omega,\end{array}\label{KKT_WF}\end{equation}
 where $a\perp b$ means the two scalars (or vectors) $a$ and $b$
are orthogonal. Observe that each $\lambda_{q}>0;$ otherwise complementarity
yields $\, p_{q}(k)=0$ for all $k\in\mathcal{N}$, which contradicts
the rate constraints.

Eliminating the multipliers $\{\mu_{q,k}\}$ corresponding to the
nonnegativity constraints and making some obvious scaling, the KKT
conditions in (\ref{KKT_WF}) are equivalent to the following MNCP:\begin{equation}
\begin{array}{lll}
0\,\leq\, p_{q}(k) & \perp & \sigma_{q}^{2}(k)+{\displaystyle \sum\limits _{r=1}^{Q}}\,|H_{qr}(k)|^{2}p_{r}(k)-|H_{qq}(k)|^{2}\,\lambda_{q}\,\geq\,0,\quad\forall k\in\mathcal{N}\text{, }\forall q\in\Omega.\\[5pt]
0\,\leq\,\lambda_{q}, &  & {\displaystyle \sum\limits _{k=1}^{N}}\,\log\left(1+\dfrac{|H_{qq}(k)|^{2}p_{q}(k)}{\sigma_{q}^{2}(k)+\sum_{r\neq q}|H_{qr}(k)|^{2}p_{r}(k)}\right)\,=\, R_{q}^{\star},\quad\forall q\in\Omega.\end{array}\label{eq:DSL achievable}\end{equation}

To proceed further we introduce an additional game, which has the
same structure of the game ${\mathscr{G}}$, 
except for the players' payoff functions, defined as in (\ref{Rate}),
but with $\sigma_{q}^{2}(k)=0$ for all $k\in\mathcal{N}$ and $q\in\Omega$.
We will refer to this game as the \emph{noiseless game}. Although
the noiseless game does not correspond to any realistic communication
system, it will be shown to be instrumental in understanding the behavior
of the original game ${\mathscr{G}}$ 
when all $\sigma_{q}^{2}(k)>0$.

Note that the conditions $\sigma_{q}^{2}(k)>0$ ensure that all the
users' rates $R_{q}(\mathbf{p}_{q},\mathbf{p}_{-q})$ in (\ref{Rate})
of the ${\mathscr{G}}$ in (\ref{Power Game}) are well-defined for
all nonnegative $\mathbf{p}\triangleq(\mathbf{p}_{q})_{q=1}^{Q}$,
with $\mathbf{p}_{q}\triangleq(p_{q}(k))_{k=1}^{N}.$ Nevertheless,
when $\sigma_{q}^{2}(k)=0$, the players' payoff functions%
\footnote{With a slight abuse of notation, we use the same symbol to denote
the payoff functions of the players in the game ${\mathscr{G}}$ in
(\ref{Power Game}) and in the noiseless game.%
} $R_{q}(\mathbf{p}_{q},\mathbf{p}_{-q})$ of the noiseless game still
remain well-defined as long as ${\sum_{r=1}^{Q}}\, p_{\, r}(k)>0$,
provided that we allow for a rate equal to $\infty$. Most importantly,
the MNCP (\ref{eq:DSL achievable}) is well defined for all \emph{nonnegative}
$\sigma_{q}^{2}(k)$, including the case when $\sigma_{q}^{2}(k)=0$
for all $k\in\mathcal{N}$ and $q\in\Omega$. The latter observation
motivates  the following definition.

\begin{definition} \label{df:noiseless equil} A set of user powers
$\mathbf{p}\triangleq(\mathbf{p}_{q})_{q=1}^{Q}$, with $\mathbf{p}_{q}\triangleq(p_{q}(k))_{k=1}^{N},$
is said to be an almost GNE of the noiseless game if there exists a set of
nonnegative scalars $\{v_{q}\}_{q=1}^{Q}$ such that \begin{equation}
0\,\leq\, p_{\, q}(k)\,\perp\,{\sum_{r=1}^{Q}}\,|H_{qr}(k)|^{2}p_{r}(k)-|H_{qq}(k)|^{2}\, v_{q}\,\geq\,0,\quad\forall k\in\mathcal{N}\text{, }\forall q\in\Omega.\label{eq:noiseless NE}\end{equation}
 We call these solutions \emph{noiseless almost equilibria,} and denote
the set of noiseless equilibria by $\mathcal{NE}_{0}$. \hfill{}$\Box$
\end{definition}


\noindent The set $\mathcal{NE}_{0}$ of users' noiseless almost equilibria constitutes
a closed, albeit not necessarily convex, cone in the space of all
users' powers. A noteworthy point about such an almost equilibrium is the
following simple property, which asserts that in the noiseless game,
every subchannel will be used by at least one user.

\begin{proposition} \label{pr:nonzero noiseless} If $\mathbf{p}\in\mathcal{NE}_{0}$,
and $\mathbf{p}\neq\mathbf{0},$ then $\sum_{r=1}^{Q}p_{\, r}(k)>0$
for all $k\in\mathcal{N}$. \end{proposition}

\begin{proof} Let ${\mathbf{p}}\in\mathcal{NE}_{0}$ be such that
$p_{\, q_{0}}(k_{0})>0$ for some pair $(q_{0},k_{0})$. By complementarity
(see (\ref{eq:noiseless NE})), we have \begin{equation}
|H_{q_{0}q_{0}}(k_{0})|^{2}\, v_{q_{0}}\,=\,{\sum_{r=1}^{Q}}\,|H_{q_{0}r}(k_{0})|^{2}\, p_{r}(k_{0})\,>\,0,\label{eq:nonzero_noiseless_1}\end{equation}
 which implies \begin{equation}
{\sum_{r=1}^{Q}}\,|H_{{q_{0}}r}(k)|^{2}\, p_{r}(k)\,\geq\,|H_{q_{0}q_{0}}(k)|^{2}\, v_{q_{0}}\,>\,0,\hspace{1.0114pc}\forall k\in\mathcal{N},\label{eq:nonzero_noiseless_2}\end{equation}
 since the $|H_{qr}(k)|^{2}$ are all positive. Equation (\ref{eq:nonzero_noiseless_2})
clearly implies: \[
{\sum_{r=1}^{Q}}\, p_{\, r}(k)\,>\,0,\hspace{1.0114pc}\forall k\in\mathcal{N},\]
 as claimed by the proposition. \vspace{0.1in}
 \end{proof}

\subsection{The noiseless asymptotic cone}

Another mathematical concept we need is that of \emph{asymptotic direction}
of a (nonconvex) set that we borrow from recession analysis \cite{ATeboulle03}.

Given the game ${\mathscr{G}}$ with rate profile ${\mathbf{R}}^{\star}\triangleq(R_{q}^{\star})_{q=1}^{Q}$,
it is possible that the sets of powers $\mathbf{p}$ that allow the
user to achieve this rate be unbounded. In essence, the asymptotic
consideration below aims at identifying such unbounded user powers.
Specifically, we consider the following noisy nonconvex level set
of users' powers corresponding to ${\mathbf{R}}^{\star}$:\begin{equation}
\mathcal{P}^{\mbox{\boldmath{$\sigma$}}}({\mathbf{R}}^{\star})\,\equiv\,\left\{ \,{\mathbf{p}}\,\geq\,0\,:\,{\sum_{k=1}^{N}}\,\log\left(1+\frac{|H_{qq}(k)|^{2}p_{q}(k)}{\sigma_{q}^{2}(k)+\sum_{r\neq q}|H_{qr}(k)|^{2}p_{r}(k)}\right)\,=\, R_{q}^{\star},\;\forall q\in\Omega\right\} ,\medskip\end{equation}

\noindent where $\mbox{\boldmath{$\sigma$}}\triangleq(\mbox{\boldmath{$\sigma$}}_{q})_{q=1}^{Q},$
with $\mbox{\boldmath{$\sigma$}}_{q}\triangleq(\sigma_{q}^{2}(k))_{k=1}^{N}.$
The \emph{asymptotic cone} of $\mathcal{P}^{\mbox{\boldmath{$\sigma$}}}({\mathbf{R}}^{\star})$,
denoted by $\mathcal{P}_{\infty}^{\mbox{\boldmath{$\sigma$}}}({\mathbf{R}}^{\star})$,
is the cone (not necessarily convex) of directions ${\mathbf{d}}\triangleq{\left(\,\mathbf{d}_{q}\,\right)_{q=1}^{Q},}$
with ${\mathbf{d}}_{q}{\triangleq\left(\, d_{q}\,(k)\right)_{k=1}^{N},}$
such that \begin{equation}
{\mathbf{d}}\,=\,{\lim_{\nu\rightarrow\infty}}\,{\dfrac{{\mathbf{p}}^{\nu}}{\tau_{\nu}}}\label{eq:def_d}\end{equation}
 for some sequence of scalars ${\left\{ \,\tau_{\nu}\,\right\} _{\nu=1}^{\infty}}$
tending to $\infty$ and some sequence of powers ${\left\{ \,{\mathbf{p}}^{\nu}\,\right\} _{\nu=1}^{\infty}}$
such that ${\mathbf{p}}^{\nu}\in P^{\mbox{\boldmath{$\sigma$}}}({\mathbf{R}}^{\star})$
for all $\nu$.

It is known \cite[Proposition 2.1.2]{ATeboulle03} that $\mathcal{P}^{\mbox{\boldmath{$\sigma$}}}({\mathbf{R}}^{\star})$
is bounded if and only if $\mathcal{P}_{\infty}^{\mbox{\boldmath{$\sigma$}}}({\mathbf{R}}^{\star})=\{\mathbf{0}\}$.
Whereas the individual sets $\mathcal{P}^{\mbox{\boldmath{$\sigma$}}}({\mathbf{R}}^{\star})$
are dependent on $\sigma_{q}^{2}(k)$, it turns out that the asymptotic
cones of all these sets are the same and equal to the following \emph{noiseless}
level set of users' powers: \[
\mathcal{P}^{0}({\mathbf{R}}^{\star})\,\equiv\,\left\{ \,{\mathbf{d}}\,{\triangleq}\,\left(\,\mathbf{d}_{\, q}\,\right)_{q=1}^{Q}\,\geq\,0\,:\,{\sum_{k:\sum_{r=1}^{Q}d_{r}(k)>0}}\,\log\left(\,1+{\frac{|H_{qq}(k)|^{2}\, d_{\, q}(k)}{{\sum_{r\neq q}}\,|H_{qr}(k)|^{2}\, d_{\, r}(k)}}\,\right)\,\leq\, R_{q}^{\star},\ \forall q\,\in\,\Omega\,\right\} ,\]
 where by convention the vacuous summation is defined to be zero (i.e.,
by definition, $\mathcal{P}^{0}({\mathbf{R}}^{\star})$ contains the
origin). The claim about the equality of $\mathcal{P}_{\infty}^{\mbox{\boldmath{$\sigma$}}}({\mathbf{R}}^{^{\star}})$
for a fixed ${\mathbf{R}}^{^{\star}}$ is formally stated and proved
in the result below.

\begin{proposition} \label{prop:3} \label{pr:inclusion} For any
$\mbox{\boldmath{$\sigma$}}>\mathbf{0}$, $\mathcal{P}_{\infty}^{\mbox{\boldmath{$\sigma$}}}({\mathbf{R}}^{^{\star}})=\mathcal{P}^{0}({\mathbf{R}}^{^{\star}})$.
\end{proposition}

\begin{proof} Let $\mbox{\boldmath{$\sigma$}}>\mathbf{0}$ be arbitrary.
Let ${\mathbf{d}}\in\mathcal{P}_{\infty}^{\mbox{\boldmath{$\sigma$}}}({\mathbf{R}}^{^{\star}})$
and $\{({\mathbf{p}}^{\nu},\tau_{\nu})\}$ be a sequence satisfying
the definition of ${\mathbf{d}}$ in (\ref{eq:def_d}). For each $q\in\,\Omega$
and all $\nu$, we have \begin{equation}
\begin{array}{lll}
R_{q}^{\star} & \geq & {\displaystyle \sum\limits _{k=1}^{N}}\,\log\left(\,1+{\dfrac{|H_{qq}(k)|^{2}\, p_{q}^{\nu}(k)}{\sigma_{q}^{2}(k)+{\displaystyle \sum_{r\neq q}}\,|H_{qr}(k)|^{2}\, p_{r}^{\nu}(k)}}\,\right)\\[0.5in]
 & \geq & {\displaystyle \sum\limits _{k:\sum_{r=1}^{Q}d_{r}(k)>0}}\,\log\left(\,1+{\dfrac{|H_{qq}(k)|^{2}\, p_{q}^{\nu}(k)}{\sigma_{q}^{2}(k)+{\displaystyle \sum_{r\neq q}}\,|H_{qr}(k)|^{2}\, p_{r}^{\nu}(k)}}\,\right)\\[0.5in]
 & = & {\displaystyle \sum\limits _{k:\sum_{r=1}^{Q}d_{r}(k)>0}}\log\left(\,1+{\dfrac{|H_{qq}(k)|^{2}\,{\dfrac{p_{q}^{\nu}(k)}{\tau_{\nu}}}}{{\dfrac{\sigma_{q}^{2}(k)}{\tau_{\nu}}}+{\displaystyle \sum_{r\neq q}}|H_{qr}(k)|^{2}{\dfrac{p_{r}^{\nu}(k)}{\tau_{\nu}}}}}\,\right).\end{array}\end{equation}
 Taking the limit $\nu\rightarrow\infty$ establishes the inclusion
$\mathcal{P}_{\infty}^{\mbox{\boldmath{$\sigma$}}}({\mathbf{R}}^{\star})\subseteq\mathcal{P}^{0}({\mathbf{R}}^{\star})$.

Conversely, it is clear that $\mathbf{0}\in\mathcal{P}^{\mbox{\boldmath{$\sigma$}}}({\mathbf{R}}^{\star})$.
Let ${\mathbf{d}}$ be a nonzero vector in $\mathcal{P}^{0}({\mathbf{R}}^{\star})$.
For any scalar $\theta>0$, we have, for all $q\in\Omega,$ \begin{equation}
\begin{array}{l}
{\displaystyle \sum\limits _{k=1}^{N}}\,\log\left(\,1+{\dfrac{\theta\,|H_{qq}(k)|^{2}\,\, d_{\, q}(k)}{\sigma_{q}^{2}(k)+\theta\,{\sum_{r\neq q}}\,|H_{qr}(k)|^{2}\, d_{\, r}(k)}}\,\right)\medskip\\
\hspace{5cm}={\displaystyle \sum\limits _{k:\sum_{r=1}^{Q}d_{r}(k)>0}}\log\left(\,1+{\dfrac{\theta\,|H_{qq}(k)|^{2}\, d_{\, q}(k)}{\sigma_{q}^{2}(k)+\theta\,{\sum_{r\neq q}}\,|H_{qr}(k)|^{2}\, d_{\, r}(k)}}\,\right)\medskip\\
\hspace{5cm}\leq{\displaystyle \sum\limits _{k:\sum_{r=1}^{Q}d_{r}(k)>0}}\,\log\left(\,1+{\dfrac{|H_{qq}(k)|^{2}\, d_{\, q}(k)}{{\sum_{r\neq q}}\,|H_{qr}(k)|^{2}\, d_{\, r}(k)}}\,\right).\end{array}\end{equation}
 Hence $\theta{\mathbf{d}}\in\mathcal{P}^{\mbox{\boldmath{$\sigma$}}}({\mathbf{R}}^{\star})$
for all $\theta>0$. It follows that ${\mathbf{d}}\in\mathcal{P}_{\infty}^{\mbox{\boldmath{$\sigma$}}}({\mathbf{R}}^{\star})$.
\end{proof}

We are now ready to introduce the key object in our proof of Theorem
\ref{th:main existence}, the cone $\mathcal{NE}_{0}\cap\mathcal{P}^{0}({\mathbf{R}}^{\mathbf{\star}})$,
which by Proposition~\ref{prop:3}, is equal to $\{\mathbf{0}\}\cup\widehat{\mathcal{NE}}_{0}({\mathbf{R}}^{\star})$,
where \begin{equation}
\begin{array}{lll}
\widehat{\mathcal{NE}}_{0}({\mathbf{R}}^{\star}) & \triangleq & \left\{ \,{\mathbf{d}}\,\in\,\mathcal{NE}_{0}\,\setminus\,\{\,\mathbf{0}\,\}\,:\,{\displaystyle \sum\limits _{k=1}^{N}}\,\log\left(\,1+{\dfrac{|H_{qq}(k)|^{2}\, d_{\, q}(k)}{{\sum_{r\neq q}}\,|H_{qr}(k)|^{2}\, d_{\, r}(k)}}\,\right)\,\leq\, R_{q}^{\star},\ \forall q\,\in\Omega\right\} \\[0.4in]
 & = & \mathcal{NE}_{0}\,\cap\,\mathcal{P}^{0}({\mathbf{R}}^{\star})\,\setminus\,\{\,\mathbf{0}\,\}.\end{array}\label{eq:def:NE_hat}\end{equation}
 Notice that ${\mathbf{d}}\in\widehat{\mathcal{NE}}_{0}({\mathbf{R}}^{\star})$
implies ${\sum_{r\neq q}|H_{qr}(k)|^{2}d_{\, r}(k)>0}$ for all $q\,\in\Omega$
and all $k\,\in\mathcal{N}$. 

\subsection{Existence results}

With the above preparation, we are ready to present our main existence
theorem. The emptiness of the set $\widehat{\mathcal{NE}}_{0}({\mathbf{R}}^{\star})$
defined in (\ref{eq:def:NE_hat}) turns out to provide a sufficient
condition for the MNCP (\ref{eq:DSL achievable}) to have a nonempty
bounded solution set.

\begin{theorem} \label{th:main existence} Given the game ${\mathscr{G}}$
with rate profile $\mathbf{R}^{\star}\triangleq(R_{q}^{\star})_{q=1}^{Q}>\mathbf{0}$,
if $\widehat{\mathcal{NE}}_{0}(\mathbf{R}^{\star})=\emptyset$, then
the game has a nonempty and bounded solution set, for all $\mbox{\boldmath{$\sigma$}}>\mathbf{0}$.
\end{theorem}

\begin{proof} We first note that the KKT conditions of the Nash problem
defined in (\ref{Power Game}) are equivalent to the following nonlinear
complementarity problem (NCP) (see (\ref{KKT_WF}) and comments thereafter)
\begin{equation}
\begin{array}{lll}
0\,\leq\, p_{\, q}(k) & \perp & \sigma_{q}^{2}(k)+{\displaystyle \sum\limits _{r=1}^{Q}}\,|H_{qr}(k)|^{2}\, p_{\, r}(k)-|H_{qq}(k)|^{2}\,\lambda_{q}\,\geq\,0,\quad\forall k\in\mathcal{N}\text{, }\forall q\in\Omega,\bigskip\\[5pt]
0\,\leq\,\lambda_{q} & \perp & {\displaystyle \sum\limits _{k=1}^{N}}\,\log\left(\,1+{\dfrac{|H_{qq}(k)|^{2}\, p_{\, q}(k)}{\sigma_{q}^{2}(k)+{\sum_{r\neq q}}\,|H_{qr}(k)|^{2}\, p_{\, r}(k)}}\,\right)-R_{q}^{\star}\,\geq\,0,\quad\forall q\in\Omega,\end{array}\label{eq:DSL achievable NCP}\end{equation}
 In turn, to show that (\ref{eq:DSL achievable NCP}) has a solution,
it suffices to prove that the solutions of the augmented NCP \[
\begin{array}{lll}
0\,\leq\, p_{\, q}(k) & \perp & \sigma_{q}^{2}(k)+{\displaystyle \sum\limits _{r=1}^{Q}}\,|H_{qr}(k)|^{2}\, p_{\, r}(k)-|H_{qq}(k)|^{2}\lambda_{q}+\tau\, p_{\, q}(k)\,\geq\,0,\quad\forall k\in\mathcal{N}\text{, }\forall q\in\Omega,\bigskip\\[5pt]
0\,\leq\,\lambda_{q} & \perp & {\displaystyle \sum\limits _{k=1}^{N}}\,\log\left(\,1+{\dfrac{|H_{qq}(k)|^{2}p_{\, q}(k)}{\sigma_{q}^{2}(k)+{\displaystyle {\sum_{r\neq q}}\,|H_{qr}(k)|^{2}p_{\, r}(k)}}}\,\right)-R_{q}^{\star}+\tau\,\lambda_{q}\,\geq\,0,\quad\forall q\in\Omega,\end{array}\]
 for all $\tau>0$ are bounded \cite[Theorem~$2.6.1$]{FPang03}.

We show the latter boundedness property by contradiction. Assume that
for some sequence of positive scalars $\{\tau_{\nu}\}$, a sequence
of solutions $\{({\mathbf{p}}^{\nu},\tau_{\nu})\}$ exists such that
each pair $({\mathbf{p}}^{\nu},\tau_{\nu})$ satisfies: \begin{equation}
\begin{array}{lll}
0\,\leq\, p_{q}^{\nu}(k) & \perp & \sigma_{q}^{2}(k)+{\displaystyle \sum\limits _{r=1}^{Q}}\,|H_{qr}(k)|^{2}p_{r}^{\nu}(k)-|H_{qq}(k)|^{2}\,\lambda_{q}^{\nu}+\tau_{\nu}\, p_{q}^{\nu}(k)\,\geq\,0\quad\forall k\in\mathcal{N}\text{, }\forall q\in\Omega,\bigskip\\[5pt]
0\,\leq\,\lambda_{q}^{\nu} & \perp & {\displaystyle {\displaystyle \sum\limits _{k=1}^{N}}\,\log\left(\,1+{\dfrac{|H_{qq}(k)|^{2}\,\, p_{q}^{\nu}(k)}{\sigma_{q}^{2}(k)+{\sum_{r\neq q}}\,|H_{qr}(k)|^{2}\,\, p_{r}^{\nu}(k)}}\,\right)-R_{q}^{\star}+\tau_{\nu}\,\lambda_{q}^{\nu}\,\geq\,0,\quad\forall q\in\Omega,}\end{array}\label{eq:aug NCP nu}\end{equation}
 and that \begin{equation}
\lim_{\nu\rightarrow\infty}\,\left[\,\Vert\,{\mathbf{p}}^{\nu}\,\Vert+{\sum_{q=1}^{Q}}\,\lambda_{q}^{\nu}\,\right]\,=\,\infty.\label{eq:unbounded}\end{equation}

From (\ref{eq:aug NCP nu}), it is clear that $\lambda_{q}^{\nu}>0$
for all $\nu$ and $q$. In fact, if a $\lambda_{q}^{\nu}=0$, then
by the first complementarity condition, $p_{q}^{\nu}(k)=0$ for all
$k$, which is not possible by the last inequality in (\ref{eq:aug NCP nu}).

Thus, it follows from the second complementarity condition that \begin{equation}
{\sum_{k=1}^{N}}\,\log\left(\,1+{\frac{|H_{qq}(k)|^{2}\, p_{q}^{\nu}(k)}{\sigma_{q}^{2}(k)+{\displaystyle \sum_{r\neq q}}\,|H_{qr}(k)|^{2}p_{r}^{\nu}(k)}}\,\right)-R_{q}^{\star}+\tau_{\nu}\,\lambda_{q}^{\nu}\,=\,0,\label{eq:binding lambda}\end{equation}
 which implies that the sequence $\{\tau_{\nu}\lambda_{q}^{\nu}\}$
is bounded for each $q\in\Omega$. We claim that ${\lim_{\nu\rightarrow\infty}\tau_{\nu}=0}$.
Otherwise, for some subsequence $\{\tau_{\nu}:\nu\in\kappa\}$, where
$\kappa$ is an infinite index set, we have ${\liminf_{\nu(\in\kappa)\rightarrow\infty}\tau_{\nu}>0}$.
Thus, the subsequence $\{\lambda_{q}^{\nu}:\nu\in\kappa\}$ is bounded
for all $q\in\Omega$. The first complementarity condition in (\ref{eq:aug NCP nu})
then implies that $\{p_{q}^{\nu}(k):\nu\in\kappa\}$ is bounded for
all $q\in\Omega$ and $k\in\mathcal{N}$. This is a contradiction
to (\ref{eq:unbounded}). Therefore, the sequence $\{\tau_{\nu}\}\downarrow0$.

Consider now the normalized sequence $\{{\mathbf{p}}^{\nu}/\Vert{\mathbf{p}}^{\nu}\Vert\}$,
which must have at least one accumulation point; moreover, any such
point must be nonzero. Let ${\mathbf{d}}^{\infty}$ be any such point.
It is not difficult to show that ${\mathbf{d}}^{\infty}$ is a nonzero almost
noiseless equilibrium. Moreover, from the inequality: \begin{equation}
{\sum_{k=1}^{N}}\,\log\left(\,1+{\frac{|H_{qq}(k)|^{2}\, p_{q}^{\nu}(k)}{\sigma_{q}^{2}(k)+{\sum_{r\neq q}}\,|H_{qr}(k)|^{2}\, p_{r}^{\nu}(k)}}\,\right)-R_{q}^{\star}\,\leq\,0,\end{equation}
 which is implied by (\ref{eq:binding lambda}), it is equally easy
to show that ${\mathbf{d}}^{\infty}\in\mathcal{P}^{0}({\mathbf{R}}^{\star})$.
Therefore, ${\mathbf{d}}^{\infty}$ is an element of $\widehat{\mathcal{NE}}_{0}({\mathbf{R}}^{\star})$,
which is a contradiction. This completes the proof of the existence
of a solution to the problem (\ref{eq:DSL achievable}). The boundedness
of such solutions can be proved in a similar way via contradiction
and by the same normalization argument. The details are not repeated.
\end{proof}

\bigskip{}

Roughly speaking, the key condition $\widehat{\mathcal{NE}}_{0}({\mathbf{R}}^{\star})=\emptyset$
in the previous theorem is just the mathematical requirement that
if the power ${\mathbf{p}}$ goes to infinity staying feasible, the
system cannot approach a (noiseless) equilibrium. As such the previous
theorem is rather natural, although it does not provide an effective
way of checking the existence and boundedness of the solutions. To
this end, however, we can now easily derive Theorem \ref{th:existence(main_body)}
from Theorem \ref{th:main existence}.

\subsection{Proof of Theorem \ref{th:existence(main_body)}}

In order to prove Theorem \ref{th:existence(main_body)} we introduce
a simple polyhedral set that will turn out to be a subset of $P^{\mbox{\boldmath{$\sigma$}}}({\mathbf{R^{\star}}})$.
\begin{equation}
{\normalsize P}_{\infty}({\mathbf{R^{\star}}})\,\triangleq\,{\prod_{k=1}^{N}}\,\left\{ \,\mathbf{r}(k)\,\in\mathbb{R}_{+}^{\, Q}\,:\,\mathbf{Z}_{k}({\mathbf{R^{\star}}})\mathbf{r}(k)\,\leq\,\mathbf{0}\,\right\} ,\end{equation}
 which is independent of the $\mbox{\boldmath{$\sigma$}}$ and where,
we recall, the matrices $\mathbf{Z}_{k}({\mathbf{R^{\star}}})$ are
defined by (\ref{Z_kL}). The key property for the existence Theorem
\ref{th:existence(main_body)} is stated in the following proposition.

\begin{proposition} \label{pr:inclusion in poly} $\widehat{\mathcal{NE}}_{0}(\mathbf{R^{\star}})\subseteq P_{\infty}(\mathbf{R^{\star}})\cap(\mathcal{NE}_{0}\setminus\{\mathbf{0}\})$.
\end{proposition}

\begin{proof} It suffices to note the following string of implications:
\begin{equation}
\begin{array}{lll}
{\mathbf{p}}\,\in\,\widehat{\mathcal{NE}}_{0}({\mathbf{R^{\star}}}) & \Rightarrow & \log\left(\,1+{\dfrac{|H_{qq}(k)|^{2}\, p_{\, q}(k)}{{\sum_{r\neq q}}\,|H_{qr}(k)|^{2}\, p_{\, r}(k)}}\,\right)\,\leq\, R_{q}^{\star},\hspace{1.0114pc}\forall\, q\,\in\Omega,\\
 & \Leftrightarrow & |H_{qq}(k)|^{2}\, p_{\, q}(k)\,\leq\,(\, e^{R_{q}^{\star}}-1\,)\,{\displaystyle \sum_{r\neq q}}\,|H_{qr}(k)|^{2}\, p_{\, r}(k),\hspace{1.0152pc}\forall\, q\,\in\Omega,\\[10pt]
 & \Leftrightarrow & {\mathbf{p}}\,\in\, P_{\infty}({\mathbf{R^{\star}}}).\end{array}\end{equation}
 where the middle equivalence is by simple exponentiation. \vspace{0.1in}
 \end{proof}

It is known that, since each matrix $\mathbf{Z}_{k}({\mathbf{R}}^{\star})$
is a Z-matrix, if each matrix $\mathbf{Z}_{k}({\mathbf{R}}^{\star})$
is also a P-matrix, then a positive vector $\mathbf{s}(k)\triangleq(s_{q}(k))_{q=1}^{Q}$
exists such that $\mathbf{s}^{T}(k)\mathbf{Z}_{k}({\mathbf{R}}^{\star})>\mathbf{0}$
\cite[Theorem $3.3.4$]{CPStone92}. But this implies that $P_{\infty}({\mathbf{R}}^{\star})=\{\mathbf{0}\}$,
and thus $\widehat{\mathcal{NE}}_{0}({\mathbf{R}}^{\star})=$ $\emptyset$.
The first assertion of Theorem \ref{th:existence(main_body)} then
follows immediately from Theorem \ref{th:main existence}. It remains
to establish the bound on the $\mathbf{p}(k)=(p_{q}(k))_{q=1}^{Q}$.
This follows easily from the following two facts: 1) any solution
${\mathbf{p}}$ of (\ref{eq:DSL achievable}) must belong to the set
$P^{\mbox{\boldmath{$\sigma$}}}({\mathbf{R}}^{\star})$; 2) a Z-matrix
that is also a P-matrix must have a nonnegative inverse. 
\hspace{\fill}\rule{1.5ex}{1.5ex}

\section{Proof of Corollary \ref{Corollary:SF_Existence_Z_maxL}}

\label{proof_Corollary:SF_Existence_Z_maxL}

Consider the matrix ${\mathbf{Z}}^{\max}({\mathbf{R^{\star}}})$ defined
by (\ref{Z_max}) in Corollary \ref{Corollary:SF_Existence_Z_maxL},
and assume it is a P-matrix. Set $\mathbf{D}(k)\triangleq{\mbox{ Diag}\left(|H_{qq}(k)|^{2}\right)_{q=1}^{Q}}$.
The first assertion in the Corollary is immediate because \begin{equation}
\mathbf{Z}_{k}({\mathbf{R}}^{\star})\geq{\mathbf{Z}}^{\max}({\mathbf{R}}^{\star})\mathbf{D}(k),\quad\forall k\in\mathcal{N}\text{,}\label{eq:PZinequality}\end{equation}
 where the inequality is intended component-wise. Therefore, since
all the matrices involved are Z-matrices, from the assumption on ${\mathbf{Z}}^{\max}({\mathbf{R}}^{\star}),$
it follows that all the $\mathbf{Z}_{k}({\mathbf{R}}^{\star})$ are
also P-matrices.\footnote{The last statement can be easily  proved using \cite[Lemma 5.3.14]{CPStone92}.
}
Lets now prove the bounds (\ref{eq:FFbounds}). Note first that \begin{equation}
\left(\begin{array}{c}
\sigma_{1}^{2}(k)\,(\, e^{R_{1}^{\ast}}-1\,)\\[5pt]
\vdots\\[5pt]
\sigma_{Q}^{2}(k)\,(\, e^{R_{Q}^{\ast}}-1\,)\end{array}\right)\,\leq\,\left[\,{\max_{r\in\Omega}}\,\sigma_{r}^{2}(k)\,\right]\,(\, e^{\mathbf{R}{\mathbf{^{\ast}}}}-\mbox{\boldmath{$1$}}\,),\quad\forall k\in\mathcal{N}\text{.}\end{equation}
 Furthermore we recall that, by (\ref{eq:PZinequality}), 
we have $\left[\mathbf{Z}_{k}({\mathbf{R}}^{\star})\right]^{-1}\leq\left[{\mathbf{Z}}^{\max}({\mathbf{R}}^{\star})\mathbf{D}(k)\right]^{-1}$ \cite{CPStone92},
and also that the inverse of a matrix that is P and Z is nonnegative
\cite[Theorem $3.11.10$]{CPStone92}. From all these facts, and recalling
(\ref{Upper_bound_GNE}), the following chain of inequalities easily
follows for every $k\in\mathcal{N}$:\begin{align}
\mathbf{D}(k)\,\left(\begin{array}{c}
\overline{p}_{1}(k)\\[5pt]
\vdots\\[5pt]
\overline{p}_{Q}(k)\end{array}\right) & \leq\mathbf{D}(k)\,[\,\mathbf{Z}_{k}({\mathbf{R}}^{\star})\,]^{-1}\left(\begin{array}{c}
\sigma_{1}^{2}(k)\,(\, e^{R_{1}^{\star}}-1\,)\\[5pt]
\vdots\\[5pt]
\sigma_{Q}^{2}(k)\,(\, e^{R_{Q}^{\star}}-1\,)\end{array}\right)\nonumber \\[1em]
 & \leq\left[{\max_{r\in\Omega}}\,\sigma_{r}^{2}(k)\,\right]\,\left[{\mathbf{Z}}^{\max}({\mathbf{R}}^{\star})\right]^{-1}\,(\, e^{\mathbf{R}{\mathbf{^{\star}}}}-\mbox{\boldmath{$1$}}\,),\end{align}
which provides the desired bound (\ref{eq:FFbounds}). \hspace{\fill
}\rule{1.5ex}{1.5ex}

\section{Proof of Proposition \ref{pr:1-tone problem}}

\label{proof_Proposition_G_one_tone}

We prove that the following three statements are equivalent for game
$\mathscr{G}$ when $N=1$:

\begin{description}
\item [{{(a)}}] The game has a nonempty solution set; 
\item [{{(b)}}] The matrix $\mathbf{Z}(\mathbf{R}^{\star})$ is a P-matrix; 
\item [{{(c)}}] $\widehat{\mathcal{NE}}_{0}(\mathbf{R}^{\star})=\emptyset$. 
\end{description}
(a) $\Rightarrow$ (b). 
 Since $N=1$, any solution of (\ref{eq:DSL achievable}) satisfies
the equations \[
\log\left(\,1+{\dfrac{|H_{qq}(k)|^{2}p_{\, q}(k)}{\sigma_{q}^{2}(k)+{\displaystyle {\sum_{r\neq q}}\,|H_{qr}(k)|^{2}p_{\, r}(k)}}}\,\right)\,=\, R_{q}^{\star},\quad\forall q\in\Omega,\]
which are easily seen to be equivalent to (\ref{eq:1-tone}). Since
the right-hand constants in (\ref{eq:1-tone}) are positive, it follows
that a vector $\mathbf{r}^{\star}\triangleq(r_{q}^{\star})_{q=1}^{Q}\geq\mathbf{0}$,
which is the solution of (\ref{eq:DSL achievable}) with $N=1$, exists
satisfying $\mathbf{Z}(\mathbf{R}^{\star})\mathbf{r}^{\star}>0$.
Hence (b) follows \cite[Theorem 3.3.4]{CPStone92}, \cite[Theorem 6.2.3]{BPlemmons79}. The implications (b) $\Rightarrow$
(c) $\Rightarrow$ (a) come directly from the more general case $N\geq1$.
Hence (a), (b), and (c) are equivalent.

It remains to establish the assertion about the uniqueness of the
solution. But this is clear because any one of the three statements
(a), (b), or (c) implies that the matrix $\mathbf{Z}(\mathbf{R}^{\star})$
is a P-matrix, thus nonsingular \cite[Theorem 6.2.3]{BPlemmons79}; and hence the system of linear equations
(\ref{eq:1-tone}) has a unique solution. \hspace{\fill}\rule{1.5ex}{1.5ex}

\section{Proof of Theorem \ref{th:uniqueness_(main body)}}
\label{proof_th:uniqueness_(main body)}

The study of the uniqueness of the solution of the GNEP is complicated
by the presence of a coupling among the feasible strategy sets of
the users, due to the rate constraints. To overcome this difficulty
we first introduce a change of variables of the game ${\mathscr{G}}$,
from the power variables $p_{q}(k)$ to a set of rate variables, in
order to obtain an equivalent formulation of the original generalized
Nash problem as a Variational Inequality (VI) problem, defined on
the Cartesian product of the users' rate admissible sets. Then, building
on this VI formulation, we derive sufficient conditions for the uniqueness
of the GNE of the original game. It is important to remark that our
VI formulation of the game ${\mathscr{G}}$ 
differs from that of \cite{LPang06}. In fact, in \cite{LPang06}
the rate maximization game was formulated as an Affine VI defined
on the Cartesian product of the users' power sets \cite[Proposition $2$]{LPang06}.
Our VI instead, is defined by a nonlinear function, which significantly
complicates the uniqueness analysis, as detailed next.

\subsection{VI formulation}

Hereafter we assume that conditions of Theorem \ref{th:existence(main_body)}
are satisfied, so that a GNE of the game ${\mathscr{G}}$ is guaranteed
to exist.

Given the game ${\mathscr{G}}$ 
we introduce the following change of variables: \begin{equation}
r_{q}(k)\,\triangleq\,\log\left(\,1+{\frac{|H_{qq}(k)|^{2}\, p_{q}(k)}{\sigma_{q}^{2}(k)+{\sum_{r\neq q}}\,|H_{qr}(k)|^{2}\, p_{r}(k)}}\,\right),\hspace{1.0228pc}k\in\mathcal{N}\text{, }q\in\Omega,\label{eq:r_k_i_def}\end{equation}
 with $r_{q}(k)$ satisfying the constraints\begin{equation}
r_{q}(k)\geq0,\text{ \ }\forall k\in\mathcal{N}\text{, }\forall q\in\Omega,\quad\text{and}\quad{\sum_{k=1}^{N}r_{q}(k)=R_{q}^{\star}},\text{ \ }\forall q\in\Omega.\label{r_k_i_constr}\end{equation}
 Observe that each $r_{q}(k)=0$ if and only if $p_{q}(k)=0.$ Given
$\mathbf{r}(k)\triangleq(r_{q}(k))_{q=1}^{Q},$ let us define the
Z-matrix $\mathbf{Z}_{k}(\mathbf{r}(k))\in\mathbb{R}^{Q\times Q}$
as$:$ \begin{equation}
\mathbf{Z}_{k}(\mathbf{r}(k))\,\triangleq\,\,\left[\begin{array}{cccc}
|H_{11}(k)|^{2} & -(e^{r_{1}(k)}-1)\,|H_{12}(k)|^{2} & \cdots & -(e^{r_{1}(k)}-1)\,|H_{1Q}(k)|^{2}\\[7pt]
-(e^{r_{2}(k)}-1)\,|H_{21}(k)|^{2} & |H_{22}(k)|^{2} & \cdots & -(e^{r_{2}(k)}-1)\,|H_{2Q}(k)|^{2}\\[7pt]
\vdots & \vdots & \ddots & \vdots\\[7pt]
-(e^{r_{Q}(k)}-1)\,|H_{Q1}(k)|^{2} & -(e^{r_{Q}(k)}-1)\,|H_{Q2}(k)|^{2} & \cdots & |H_{QQ}(k)|^{2}\end{array}\right]\,.\label{Z_k_r_k}\end{equation}
 From (\ref{r_k_i_constr}), we have $\mathbf{Z}_{k}(\mathbf{r}(k))\,\geq\mathbf{Z}_{k}(\mathbf{R}^{\star})\,$\ for
all $k\in\mathcal{N}$, where $\mathbf{Z}_{k}(\mathbf{R}^{\star})$
is defined in (\ref{Z_kL}). It follows that each $\mathbf{Z}_{k}(\mathbf{r}(k))$
is a P-matrix.

According to (\ref{eq:r_k_i_def}), the users' powers $\mathbf{p}(k)\triangleq(p_{q}(k))_{q=1}^{Q}$
are related to the rates $\mathbf{r}(k)$ by the following function
\begin{equation}
\left(\begin{array}{c}
p_{1}(k)\\[3pt]
\vdots\\[3pt]
p_{Q}(k)\end{array}\right)\,=\mathbf{\boldsymbol{\phi}}(k,\mathbf{r}(k))\triangleq\,\left(\,\mathbf{Z}_{k}(\mathbf{r}(k))\right)^{-1}\left(\begin{array}{c}
\sigma_{1}^{2}(k)\,(\, e^{r_{1}(k)}-1\,)\\[3pt]
\vdots\\[3pt]
\sigma_{Q}^{2}(k)\,(\, e^{r_{Q}(k)}-1\,)\end{array}\right)\,,\quad k\in\mathcal{N}\text{.}\label{eq:def_phi_k}\end{equation}

\noindent Observe that $\left(\,\mathbf{Z}_{k}(\mathbf{r}(k))\right)^{-1}$
is well-defined, since $\,\mathbf{Z}_{k}(\mathbf{r}(k))$ is a P-matrix \cite[Theorem 3.11.10]{CPStone92}.

Using (\ref{r_k_i_constr}) and (\ref{eq:def_phi_k}), and the fact
that each $p_{q}(k)=0$ if and only if $r_{q}(k)=0$, the KKT conditions
of the Nash problem (\ref{Power Game}) can be rewritten as (see (\ref{eq:DSL achievable})):
\begin{equation}
\begin{array}{lll}
0\,\leq\, r_{q}(k) & \perp & \sigma_{q}^{2}(k)+{\sum\limits _{r=1}^{Q}}\,|H_{qr}(k)|^{2}\,\phi_{r}(k,\mathbf{r}(k))-|H_{qq}(k)|^{2}\,\lambda_{q}\,\geq\,0,\quad\forall k\in\mathcal{N}\text{, }\forall q\in\Omega\\[5pt]
0\,\leq\,\lambda_{q}, &  & {\displaystyle \sum\limits _{k=1}^{N}}\, r_{q}(k)\,={R_{q}^{\star}},\quad\forall q\in\Omega\end{array}\label{eq:DSL achievable in rates}\end{equation}
 where $\phi_{r}(k,\mathbf{r}(k))$ denotes the $r$-th component
of $\mathbf{\boldsymbol{\phi}}(k,\mathbf{r}(k)),$ defined in (\ref{eq:def_phi_k}).
It is easy to see that (\ref{eq:DSL achievable in rates}) is equivalent
to (note that as usual $\lambda_{q}>0$ for any solution of (\ref{eq:DSL achievable in rates})),
$\forall k\in\mathcal{N}\text{, }\forall q\in\Omega$, \begin{equation}
\begin{array}{lll}
0\,\leq\, r_{q}(k) & \perp & \log\left(\sigma_{q}^{2}(k)+{\sum\limits _{r=1}^{Q}}\,|H_{qr}(k)|^{2}\,\phi_{r}(k,\mathbf{r}(k))\right)-\log\left(|H_{qq}(k)|^{2}\right)+\nu_{q}\,\geq\,0,\\[5pt]
\nu_{q}\text{ free,} &  & {\displaystyle \sum\limits _{k=1}^{N}}\, r_{q}(k)\,={R_{q}^{\star}},\end{array}\label{KKT_VI}\end{equation}
 Let us define\begin{equation}
\begin{array}{l}
\Phi_{q}(k,\mathbf{r}(k))\triangleq\log\left(\sigma_{q}^{2}(k)+{\sum\limits _{r=1}^{Q}}\,|H_{qr}(k)|^{2}\,\phi_{r}(k,\mathbf{r}(k))\right)-\log\left(|H_{qq}(k)|^{2}\right),\quad k\in\mathcal{N}\text{, }q\in\Omega.\medskip\\
\mathbf{\Phi}_{q}({\mathbf{r}})\triangleq(\Phi_{q}(k,\mathbf{r}(k)))_{k=1}^{N},\quad\mathbf{\Phi}({\mathbf{r}})\triangleq(\mathbf{\Phi}_{q}({\mathbf{r}}))_{q=1}^{Q},\quad{\mathbf{r}}\triangleq(\mathbf{r}_{q})_{q=1}^{Q},\end{array}\label{def_Phi}\end{equation}
 with $\mathbf{r}_{q}\triangleq(r_{q}(k))_{k=1}^{N}$. Observe that
each $\Phi_{q}(k,\mathbf{r}(k))$ in (\ref{def_Phi}) is a well-defined
continuously differentiable function on the rectangular box ${[\mathbf{0},\mathbf{R}^{\star}]\triangleq\prod_{q=1}^{Q}\,[0,}R_{q}^{\star}{]\subset\mathbb{R}^{Q}}$;
thus $\mathbf{\Phi}$ in (\ref{def_Phi}) is a well-defined continuously
differentiable function on ${[\mathbf{0},\mathbf{R}^{\star}]}^{N}$.

Using (\ref{def_Phi}), one can see that (\ref{KKT_VI}) are the KKT
conditions of the VI $(U,\mathbf{\Phi})$ \cite[Proposition $1.3.4$]{FPang03},
where $U$ is the Cartesian product of users' rate sets, defined as
\begin{equation}
U\,\equiv\,\prod_{q=1}^{Q}\, U_{q},\hspace{1.0266pc}\mbox{where}\hspace{1.0266pc}U_{q}\,\triangleq\,\left\{ \,\mathbf{r}_{q}\,\in\,{\mathbb{R}_{+}^{N}}\,:{\sum_{k=1}^{N}}\, r_{q}(k)\,=R_{q}^{\star}\,\right\} ,\label{eq:VI-def}\end{equation}
 and $\mathbf{\Phi}$ is the continuously differentiable function
on ${[\mathbf{0},\mathbf{R}^{\star}]}^{N}$ defined in (\ref{def_Phi}).

By definition, it follows that a tuple $\mathbf{r}^{\star}\triangleq(\mathbf{r}_{q}^{\star})_{q=1}^{Q}$
is a solution of the VI$(U,\mathbf{\Phi})$ defined above if and only
if, for all $\mathbf{r}_{q}\in U_{q}$ and $q\in\Omega,$\begin{equation}
{\sum_{k=1}^{N}}\,\left(\, r_{q}(k)-r_{q}^{\star}(k)\,\right)\,\left[\,\log\left(\sigma_{q}^{2}(k)+{\sum\limits _{r=1}^{Q}}\,|H_{qr}(k)|^{2}\,\phi_{r}(k,\mathbf{r}_{k}^{\star})\right)-\log\left(|H_{qq}(k)|^{2}\,\right)\,\right]\,\geq\,0.\label{eq:VI_1}\end{equation}

We rewrite now condition (\ref{eq:VI_1}) in a more useful form. To
this end, let introduce$\,$\ \begin{equation}
\tau_{q}(k,\mathbf{p}(k))\,\triangleq\,\sigma_{q}^{2}(k)+{\sum_{r\neq q}}\,|H_{qr}(k)|^{2}\, p_{r}(k),\quad k\in\mathcal{N}\text{, }q\in\Omega,\label{eq:def_tau_i_p_k}\end{equation}
 with $\mathbf{p}(k)\,\triangleq\,\,\left(\, p_{q}(k)\right)_{q=1}^{Q}.$
For any solution $\mathbf{p}^{\star}=(\mathbf{p}^{\star}(k))_{k=1}^{N}$
of (\ref{KKT_WF}) (i.e., any GNE of the game ${\mathscr{G}}$ we
have $\mathbf{p}^{\star}\leq\mathbf{\bar{p}}$, where $\mathbf{\bar{p}}=(\mathbf{\bar{p}}(k))_{k=1}^{N},$
with $\mathbf{\bar{p}}(k)=(\bar{p}_{r}(k))_{r=1}^{Q}$ and $\bar{p}_{r}(k)$
defined in (\ref{Upper_bound_GNE}). It follows that \begin{equation}
\sigma_{q}^{2}(k)\leq\tau_{q}(k,\mathbf{p}^{\star}(k))\,\leq\,\bar{\tau}_{q}(k)\,\triangleq\,\sigma_{q}^{2}(k)+{\sum_{r\neq q}}\,|H_{qr}(k)|^{2}\,\bar{p}_{r}(k),\quad\forall k\in\mathcal{N}\text{, }\forall q\in\Omega.\label{eq:upper_bound_tau_i_p_k}\end{equation}

Using (\ref{eq:def_tau_i_p_k}), 
we can write \begin{equation}
\log\left(\sigma_{q}^{2}(k)+{\sum\limits _{r\neq q}^{Q}}\,|H_{qr}(k)|^{2}\, p_{r}(k)+|H_{qq}(k)|^{2}\, p_{q}(k)\right)=r_{q}(k)+\log\left(\tau_{q}(k,\mathbf{p}(k))\,\right),\label{eq:phi_of_tau_i_k}\end{equation}
 so that condition (\ref{eq:VI_1}) becomes\begin{equation}
{\sum_{k=1}^{N}}\,\left(\, r_{q}(k)-r_{q}^{\star}(k)\,\right)\,\left(\,\log\left(\tau_{q}(k,\mathbf{p}^{\star}(k))\,\right)-\log\left(|H_{qq}(k)|^{2}\right)+r_{q}^{\star}(k)\,\right)\,\geq\,0,\hspace{1.0343pc}\forall\,\mathbf{r}_{q}\in U_{q},\quad\forall q\in\Omega,\label{eq:VI}\end{equation}
 where $\tau_{q}(k,\mathbf{p}^{\star}(k))$ is defined in (\ref{eq:def_tau_i_p_k})
and $\mathbf{p}^{\star}(k)\triangleq(p_{q}^{\star}(k))_{q=1}^{Q},$
with each $p_{q}^{\star}(k)=\phi_{q}(k,\mathbf{r}^{\star}(k)),$ and
$\phi_{q}(\mathbf{\cdot})$ given in (\ref{eq:def_phi_k}). Condition
(\ref{eq:VI}) will be instrumental for the study of the uniqueness
of the GNE, as shown next.

\subsection{Uniqueness analysis}

Building on (\ref{eq:VI}), we derive now sufficient conditions for
the uniqueness of the GNE of the game ${\mathscr{G}}$. 
Let $\widehat{{\mathbf{p}}}^{\,(\nu)}\triangleq(\widehat{\mathbf{p}}_{q}^{(\nu)})_{q=1}^{Q},$
for $\nu=1,2,$ be any two solutions of the Nash problem in (\ref{Power Game}).
Given $\nu=1,2$, $k\in\mathcal{N},$ $\ $and $q\in\Omega,$ let
us define \begin{equation}
\widehat{r}_{q}^{(\nu)}(k)\,\triangleq\log\left(\,1+{\frac{|H_{qq}(k)|^{2}\,\widehat{p}_{q}^{(\nu)}(k)}{\sigma_{q}^{2}(k)+{\sum_{r\neq q}}\,|H_{qr}(k)|^{2}\,\widehat{p}_{r}^{(\nu)}(k)}}\,\right),\hspace{1.0538pc}\text{and}\hspace{1.0578pc}\widehat{\tau}_{q}^{(\nu)}(k)\triangleq\tau_{q}(k,\widehat{\mathbf{p}}^{(\nu)}(k)).\quad\label{eq:def_two_sol_VI}\end{equation}
 Adding the following two inequalities, which are obtained from the
characterization (\ref{eq:VI}) of a solution to the VI $(U,\mathbf{\Phi})$:\begin{equation}
{\sum_{k=1}^{N}}\,\left(\,\widehat{r}_{q}^{\,(2)}(k)-\widehat{r}_{q}^{\,(1)}(k)\,\right)\,\left(\,-\log\left(|H_{qq}(k)|^{2}\right)+\log(\widehat{\tau}_{q}^{\,(1)}(k))+\widehat{r}_{q}^{\,(1)}(k)\,\right)\,\geq\,0,\quad\forall q\in\Omega,\end{equation}
 and \begin{equation}
{\sum_{k=1}^{N}}\,\left(\,\widehat{r}_{q}^{\,(1)}(k)-\widehat{r}_{q}^{\,(2)}(k)\,\right)\,\left(\,-\log\left(|H_{qq}(k)|^{2}\right)+\log(\widehat{\tau}_{q}^{(2)}(k))+\widehat{r}_{q}^{\,(2)}(k)\,\right)\,\geq\,0,\quad\forall q\in\Omega,\end{equation}
 and rearranging terms, we obtain \begin{align*}
{\sum_{k=1}^{N}}\, & \left(\,\widehat{r}_{q}^{\,(2)}(k)\,\,-\widehat{r}_{q}^{\,(1)}(k)\right)^{2}\,\leq\,{\sum_{k=1}^{N}}\,\left(\,\widehat{r}_{q}^{\,(2)}(k)-\widehat{r}_{q}^{\,(1)}(k)\,\right)\,\left(\,\log(\widehat{\tau}_{q}^{\,(1)}(k))-\log(\widehat{\tau}_{q}^{\,(2)}(k))\right)\\
\quad & \leq\,\sqrt{{\sum_{k=1}^{N}}\,\left(\,\widehat{r}_{q}^{\,(2)}(k)-\widehat{r}_{q}^{\,(1)}(k)\,\right)^{2}\,}\sqrt{\sum_{k=1}^{N}\left(\,\log(\widehat{\tau}_{q}^{\,(1)}(k))-\log(\widehat{\tau}_{q}^{(2)}(k))\right)^{2}}\quad\forall q\in\Omega,\end{align*}
 which implies \begin{equation}
\left\Vert \,\widehat{\mathbf{r}}_{q}^{(2)}-\widehat{\mathbf{r}}_{q}^{(1)}\,\right\Vert _{2}\,\leq\,\left\Vert \,\log(\widehat{\mathbf{\boldsymbol{\tau}}}_{q}^{(2)})-\log(\widehat{\mathbf{\boldsymbol{\tau}}}_{q}^{(1)})\,\right\Vert _{2},\quad\forall q\in\Omega,\label{eq:uie-1}\end{equation}
 where $\widehat{\mathbf{\boldsymbol{\tau}}}_{q}^{(\nu)}\triangleq(\widehat{\tau}_{q}^{(\nu)}(k))_{k=1}^{N},$
for $\nu=1,2,$ and $\log(\widehat{\mathbf{\boldsymbol{\tau}}}_{q}^{(\nu)})$
has to be intended as the vector whose $k$-th component is $\log(\widehat{\tau}_{q}^{(\nu)}(k))$.

Invoking the mean-value theorem for the logarithmic function, we have
that there exists some scalar $s_{q}(k)$ such that \begin{equation}
|H_{qq}(k)|^{2}\frac{\widehat{p}_{q}^{(2)}(k)}{\widehat{\tau}_{q}^{\,(2)}(k)}\leq s_{q}(k)\leq|H_{qq}(k)|^{2}\frac{\widehat{p}_{q}^{(1)}(k)}{\widehat{\tau}_{q}^{\,(1)}(k)},\label{ineq_on_s_k_i}\end{equation}
 and, for each $q\in\Omega$ and $k\in\mathcal{N}$,\begin{align}
\widehat{r}_{q}^{\,(2)}(k)-\widehat{r}_{q}^{\,(1)}(k) & =\log\left(\,1+{\dfrac{|H_{qq}(k)|^{2}\,\widehat{p}_{q}^{(2)}(k)}{\widehat{\tau}_{q}^{\,(2)}(k)}}\,\right)-\log\left(\,1+{\dfrac{|H_{qq}(k)|^{2}\,\widehat{p}_{q}^{(1)}(k)}{\widehat{\tau}_{q}^{\,(1)}(k)}}\,\right)\bigskip\nonumber \\
 & =\left(\,{\dfrac{\widehat{p}_{q}^{(2)}(k)}{\widehat{\tau}_{q}^{\,(2)}(k)}}-{\dfrac{\widehat{p}_{q}^{(1)}(k)}{\widehat{\tau}_{q}^{\,(1)}(k)}}\,\right){\dfrac{|H_{qq}(k)|^{2}}{1+s_{q}(k)}}\nonumber \\
 & =\left(\,{\dfrac{\widehat{p}_{q}^{(2)}(k)-\widehat{p}_{q}^{(1)}(k)}{\widehat{\tau}_{q}^{\,(2)}(k)}}\,\right)\,{\dfrac{|H_{qq}(k)|^{2}}{1+s_{q}(k)}}\,+\left({\dfrac{1}{\widehat{\tau}_{q}^{\,(2)}(k)}}-{\dfrac{1}{\widehat{\tau}_{q}^{\,(1)}(k)}}\right)\,{\dfrac{|H_{qq}(k)|^{2}\,\widehat{p}_{q}^{(1)}(k)}{1+s_{q}(k)}}\bigskip\nonumber \\
 & =\left(\,\left(\,\widehat{p}_{q}^{(2)}(k)-\widehat{p}_{q}^{(1)}(k)\,\right)+{\dfrac{\widehat{p}_{q}^{(1)}(k)}{\widehat{\tau}_{q}^{\,(1)}(k)}}\,{\displaystyle \sum_{r\neq q}}\,|H_{qr}(k)|^{2}\left(\,\widehat{p}_{r}^{(1)}(k)-\widehat{p}_{r}^{(2)}(k)\,\right)\,\right)\bigskip\nonumber \\
 & \hspace{0.6cm}\cdot{\dfrac{|H_{qq}(k)|^{2}}{\widehat{\tau}_{q}^{\,(2)}(k)\,(\,1+s_{q}(k)\,)}.}\label{eq:Delta_r_k}\end{align}
 Similarly, there exists some scalar $\omega_{q}(k)$ such that \begin{equation}
\widehat{\tau}_{q}^{\,(2)}(k)\leq\omega_{q}(k)\leq\widehat{\tau}_{q}^{\,(1)}(k),\label{ineq_on_omega}\end{equation}
 and, for each $q\in\Omega$ and $k\in\mathcal{N}$, \begin{equation}
\log(\widehat{\tau}_{q}^{\,(2)}(k))-\log(\widehat{\tau}_{q}^{\,(1)}(k))\,=\,{\frac{1}{\omega_{q}(k)}}\,(\,\widehat{\tau}_{q}^{\,(2)}(k)-\widehat{\tau}_{q}^{\,(1)}(k)\,)\,=\,{\frac{1}{\omega_{q}(k)}}\,{\sum_{r\neq q}}\,|H_{qr}(k)|^{2}\,(\,\widehat{p}_{r}^{(2)}(k)-\widehat{p}_{r}^{(1)}(k)\,).\label{eq:mean-theo_on_log_tau}\end{equation}
 Introducing \begin{equation}
\varepsilon_{q}(k)\triangleq|H_{qq}(k)|^{2}(\widehat{p}_{q}^{(2)}(k)-\widehat{p}_{q}^{(1)}(k)),\label{eq:def_e_k_i}\end{equation}
 and using (\ref{eq:Delta_r_k})\ and (\ref{eq:mean-theo_on_log_tau}),
the inequality (\ref{eq:uie-1}) becomes\begin{equation}
\begin{array}{l}
\sqrt{{\displaystyle \sum\limits _{k=1}^{N}}\,\left(\,\left(\,\varepsilon_{q}(k)+{\dfrac{|H_{qq}(k)|^{2}\,\widehat{p}_{q}^{(1)}(k)}{\widehat{\tau}_{q}^{\,(1)}(k)}}\,{\displaystyle \sum\limits _{r\neq q}}\,|H_{qr}(k)|^{2}\left(\widehat{p}_{r}^{(2)}(k)-\widehat{p}_{r}^{(1)}(k)\right)\right){\dfrac{1}{\widehat{\tau}_{q}^{\,(2)}(k)\,(\,1+s_{q}(k)\,)}}\,\right)^{2}}\bigskip\\
\leq\,\sqrt{{\displaystyle \sum\limits _{k=1}^{N}}\,\left({\dfrac{1}{\omega_{q}(k)}{\displaystyle \sum\limits _{r\neq q}}}\,|H_{qr}(k)|^{2}\,\left(\widehat{p}_{r}^{(2)}(k)-\widehat{p}_{r}^{(1)}(k)\right)\right)^{2}}.\end{array}\label{eq:uie-1_(2)}\end{equation}

By the triangle inequality and rearranging terms, from (\ref{eq:uie-1_(2)})
it follows \begin{equation}
\begin{array}{l}
\sqrt{{\displaystyle \sum\limits _{k=1}^{N}}\,\left(\,{\dfrac{\varepsilon_{q}(k)}{\widehat{\tau}_{q}^{\,(2)}(k)\,(\,1+s_{q}(k)\,)}}\,\right)^{2}}\,\leq\sqrt{{\displaystyle \sum\limits _{k=1}^{N}}\,\left({\dfrac{1}{\omega_{q}(k)}}\,{\displaystyle \sum\limits _{r\neq q}}\,|H_{qr}(k)|^{2}\,\left(\widehat{p}_{r}^{(2)}(k)-\widehat{p}_{r}^{(1)}(k)\right)\,\right)^{2}}\\[0.3in]
+\sqrt{{\displaystyle \sum\limits _{k=1}^{N}}\,\left(\,\left(\,{\dfrac{|H_{qq}(k)|^{2}\,\widehat{p}_{q}^{(1)}(k)}{\widehat{\tau}_{q}^{\,(1)}(k)}}\,{\displaystyle \sum\limits _{r\neq q}}\,|H_{qr}(k)|^{2}\,\left(\widehat{p}_{r}^{(2)}(k)-\widehat{p}_{r}^{(1)}(k)\right)\,\right){\dfrac{1}{\widehat{\tau}_{q}^{\,(2)}(k)\,(\,1+s_{q}(k)\,)}}\,\right)^{2}}.\end{array}\label{eq:uie-1_(3)}\end{equation}
 We bound now (\ref{eq:uie-1_(3)}) using the following: $\forall q\in\Omega,$
$\forall k\in\mathcal{N}$, \begin{align}
{\frac{|H_{qq}(k)|^{2}\,\widehat{p}_{q}^{(\nu)}(k)}{\widehat{\tau}_{q}^{(\nu)}(k)}} & \leq{\operatorname{e}}^{-R_{q}^{\star}}-1,\quad\nu=1,2,\label{ineq_1}\\
{\dfrac{1}{\,1+s_{q}(k)\,}} & {\geq}\text{ }{\operatorname{e}}^{-R_{q}^{\star}},\label{ineq_2}\\
\dfrac{1}{\bar{\tau}_{q}(k)\,}\leq\dfrac{1}{\widehat{\tau}_{q}^{(\nu)}(k)\,} & \leq\dfrac{1}{\sigma_{q}^{2}(k)\,},\quad\nu=1,2,\label{ineq_3}\\
\dfrac{1}{\omega_{q}(k)} & \leq\dfrac{1}{\sigma_{q}^{2}(k)},\label{ineq_4}\end{align}
 where (\ref{ineq_1}) follows from (\ref{r_k_i_constr}), (\ref{ineq_2})
from (\ref{ineq_on_s_k_i}) and (\ref{ineq_1}), (\ref{ineq_3}) from
(\ref{eq:upper_bound_tau_i_p_k}), and (\ref{ineq_4}) from (\ref{ineq_on_omega}).
Using (\ref{ineq_1})-(\ref{ineq_4}), (\ref{eq:uie-1_(3)}) can be
bound as\begin{align}
{\operatorname{e}}^{-R_{q}^{\star}}\,\sqrt{{\displaystyle \sum\limits _{k=1}^{N}}\,\left({\dfrac{\varepsilon_{q}(k)}{\bar{\tau}_{q}(k)}}\right)^{2}} & \leq\sqrt{{\displaystyle \sum\limits _{k=1}^{N}}\,\left({\dfrac{\varepsilon_{q}(k)}{\widehat{\tau}_{q}^{\,(2)}(k)\,(\,1+s_{q}(k)\,)}}\right)^{2}}\,\,\label{ineq_error_1}\\
 & \leq\sqrt{{\displaystyle \sum\limits _{k=1}^{N}}\,\left({\dfrac{{\operatorname{e}}^{R_{q}^{\star}}-1}{\sigma_{q}^{2}(k)}}\,{\displaystyle \sum\limits _{r\neq q}}\,|H_{qr}(k)|^{2}\,\left(\widehat{p}_{r}^{(2)}(k)-\widehat{p}_{r}^{(1)}(k)\right)\right)^{2}}\nonumber \\
 & \quad+\sqrt{{\displaystyle \sum\limits _{k=1}^{N}}\,\left({\dfrac{1}{\sigma_{q}^{2}(k)}}\,{\displaystyle \sum\limits _{r\neq q}}\,|H_{qr}(k)|^{2}\,\left(\widehat{p}_{r}^{(2)}(k)-\widehat{p}_{r}^{(1)}(k)\right)\right)^{2}}\label{ineq_error_2}\\
 & ={\operatorname{e}}^{R_{q}^{\star}}\,\sqrt{{\sum_{k=1}^{N}}\,\left({\frac{1}{\sigma_{q}^{2}(k)}}\,{\sum_{r\neq q}}\,|H_{qr}(k)|^{2}\,\left(\widehat{p}_{r}^{(2)}(k)-\widehat{p}_{r}^{(1)}(k)\right)\right)^{2}}\label{ineq_error_3}\\
 & ={\operatorname{e}}^{R_{q}^{\star}}\,\sqrt{{\sum_{k=1}^{N}}\,\left({\sum_{r\neq q}}\,\widehat{\beta}_{qr}(k)\left({\frac{\varepsilon_{r}(k)}{\bar{\tau}_{r}(k)}}\right)\right)^{2}}\label{ineq_error_4}\\
 & \leq{\operatorname{e}}^{R_{q}^{\star}}\,{\sum_{r\neq q}}\left(\max_{k\in\mathcal{N}}\widehat{\beta}_{qr}(k)\right)\sqrt{{\sum_{k=1}^{N}}\,\left({\frac{\varepsilon_{r}(k)}{\bar{\tau}_{r}(k)}}\right)^{2}\,},\quad\forall q\in\Omega,\label{ineq_error_5}\end{align}
 where: (\ref{ineq_error_1}) follows from (\ref{ineq_2}) and (\ref{ineq_3});
(\ref{ineq_error_2}) follows from (\ref{ineq_1}), (\ref{ineq_3})
and (\ref{ineq_4}); and in (\ref{ineq_error_4}) we have defined\begin{equation}
\widehat{\beta}_{qr}(k)\triangleq\frac{\bar{\tau}_{r}(k)}{\sigma_{q}^{2}(k)}\frac{|H_{qr}(k)|^{2}}{|H_{rr}(k)|^{2}},\quad q,r\in\Omega,\text{ and }k\in\mathcal{N}\text{.}\label{eq:def_beta}\end{equation}

Let $\mathbf{B}=\mathbf{B}(\mathbf{R}^{\star})\triangleq{\left[b_{qr}\right]_{q,r=1}^{Q}}$
be the nonnegative matrix, where \begin{equation}
b_{qr}\,\triangleq\,\left\{ \begin{array}{ll}
{\operatorname{e}}^{-R_{q}^{\star}} & \mbox{if $i\,=\, j$}\\[5pt]
{\operatorname{e}}^{R_{q}^{\star}}\,\max\limits _{k\in\mathcal{N}}\widehat{\beta}_{qr}(k) & \mbox{if $i\,\neq\, j$},\end{array}\right.\label{eq:matrix B}\end{equation}
 and let $\overline{\mathbf{B}}$ be the {}``comparison matrix\textquotedblright\ of
$\mathbf{B}$, i.e., the matrix whose diagonal entries are the same
as those of $\mathbf{B}$ and the off-diagonal entries are the negatives
of those of $\mathbf{B}$ (see (\ref{Beta_bar_matrix})). Note that
$\overline{\mathbf{B}}$ is a Z-matrix.

Introducing \begin{equation}
t_{q}\triangleq\sqrt{{\sum_{k=1}^{N}}\,\left(\,{\frac{\varepsilon_{q}(k)}{\bar{\tau}_{q}(k)}}\,\right)^{2}},\quad q\in\Omega,\label{eq:def:q_i}\end{equation}
 with $\varepsilon_{q}(k)$ defined in (\ref{eq:def_e_k_i}), and
concatenating the inequalities in (\ref{ineq_error_1}) for all $q\in\Omega,$
we deduce\begin{equation}
\overline{\mathbf{B}}\mathbf{t}\leq\mathbf{0},\label{eq:vect_ineq}\end{equation}
 with $\mathbf{t}\triangleq(t_{q})_{q=1}^{Q}$. If $\overline{\mathbf{B}}$
is a P-matrix, then it must have a nonnegative inverse \cite[Theorem 3.11.10]{CPStone92}. Thus, by (\ref{eq:vect_ineq}),
we have $\mathbf{t}\leq\mathbf{0},$ which yields $\mathbf{t}=\mathbf{0.}$
This proves the uniqueness of the GNE, under conditions of Theorem
\ref{th:uniqueness_(main body)}.

\begin{remark}\rm  An alternative approach to establish the uniqueness
of the solution of the Nash problem (\ref{Power Game}) is to show
that under a similar hypothesis, the function $\mathbf{\Phi}(\mathbf{r})$
in (\ref{def_Phi}) is a {}``uniformly P-function\textquotedblright\ on
the Cartesian product set $U$. In turn, the latter can be proved
by showing that the Jacobian matrix $\mathbf{J\Phi}(\mathbf{r})$
of the function $\mathbf{\Phi}(\mathbf{r})$ is a {}``partitioned
P-matrix\textquotedblright\ uniformly for all $\mathbf{r}\in U$.
We adopt the above proof because it can be used directly in the convergence
analysis of the distributed algorithm to be presented subsequently.
\end{remark}

\section{Proof of Corollary \ref{Corollary:SF_Uniqueness}.}

\label{proof_Corollary:SF_Uniqueness}To prove the desired sufficient
conditions for $\overline{\mathbf{B}}(\mathbf{R}^{\star})$ in (\ref{Beta_bar_matrix})
to be a P-matrix, we use the bounds (\ref{eq:FFbounds}) in Corollary
\ref{Corollary:SF_Existence_Z_maxL}, as shown next.

We provide first an upper bound of each $\widehat{\beta}_{qr}(k)$,
defined in (\ref{eq:def_beta}). Let $\overline{\mathbf{d}}\triangleq(\overline{d}_{q})_{q=1}^{Q}$
be defined as \begin{equation}
\overline{\mathbf{d}}\triangleq\left({\mathbf{Z}}^{\max}({\mathbf{R}}^{\star})\right)^{-1}\left(e^{\mathbf{R}^{\star}}-\mathbf{1}\right),\label{eq:def_d_vect}\end{equation}
 with ${\mathbf{Z}}^{\max}({\mathbf{R}}^{\star})$ given in (\ref{Z_max})
and $\mathbf{R}^{\star}=(R_{q}^{\star})_{q=1}^{Q}$. By (\ref{eq:FFbounds}),
we have \begin{equation}
\bar{\tau}_{q}(k)\,\leq\,\sigma_{q}^{2}(k)+\left(\,{\max_{r^{^{\prime}}\in\Omega}}\,\sigma_{r^{^{\prime}}}^{2}(k)\,\right)\,{\sum_{r\neq q}}\,\dfrac{\left\vert H_{qr}(k)\right\vert ^{2}}{\left\vert H_{rr}(k)\right\vert ^{2}}\overline{d}_{r},\quad\forall k\in\mathcal{N},\text{ }\forall q,r\in\Omega,\text{ and }q\neq r,\label{eq:up_bound_on_tau_bar}\end{equation}
 where $\bar{\tau}_{q}(k)$ is defined in (\ref{eq:upper_bound_tau_i_p_k}).
Introducing (\ref{eq:up_bound_on_tau_bar}) in (\ref{eq:def_beta}),
we obtain \begin{equation}
\widehat{\beta}_{qr}(k)\leq\,\dfrac{\left\vert H_{qr}(k)\right\vert ^{2}}{\left\vert H_{rr}(k)\right\vert ^{2}}\left[\,{\frac{\sigma_{r}^{2}(k)}{\sigma_{q}^{2}(k)}}+\,\left({\max_{r^{\,\prime}\in\Omega}}\,{\frac{\sigma_{r^{\,\prime}}^{2}(k)\,}{\sigma_{q}^{2}(k)}}\right){\sum_{r^{\,\prime}\neq q}}\,\dfrac{\left\vert H_{qr^{\,\prime}}(k)\right\vert ^{2}}{\left\vert H_{r^{\,\prime}r^{\,\prime}}(k)\right\vert ^{2}}\overline{d}_{r^{\,\prime}}\,\right],\quad\forall k\in\mathcal{N},\text{ }\forall q,r\in\Omega,\text{ and }q\neq r.\end{equation}
 Hence, recalling the definitions of the constants $\chi$ (assumed
in $(0,1)$) and $\rho\geq1,$ as given in (\ref{eq:def:_xi}) and
(\ref{eq:def_rho}), respectively, we deduce\begin{equation}
\begin{array}{lll}
\max\limits _{k\in\mathcal{N}}\widehat{\beta}_{qr}(k) & \leq & \left(\max\limits _{k\in\mathcal{N}}\dfrac{\left\vert H_{qr}(k)\right\vert ^{2}}{\left\vert H_{rr}(k)\right\vert ^{2}}\right)\bigskip\\
 &  & \cdot\left\{ \,\left(\,{\max\limits _{k\,\in\mathcal{N}}\dfrac{\sigma_{r}^{2}(k)}{\sigma_{q}^{2}(k)}}\,\right)+\left[\,{\max\limits _{r^{\,\prime}\in\Omega}}\,\left({\max\limits _{k\,\in\mathcal{N}}}\,{\dfrac{\sigma_{r^{\,\prime}}^{2}(k)}{\sigma_{q}^{2}(k)}}\right)\right]\,{\displaystyle \sum\limits _{r^{\,\prime}\neq q}}\,\left(\max\limits _{k\in\mathcal{N}}\dfrac{\left\vert H_{qr^{\,\prime}}(k)\right\vert ^{2}}{\left\vert H_{r^{\,\prime}r^{\,\prime}}(k)\right\vert ^{2}}\,\right)\,\,\,\widehat{d}_{r^{\,\prime}}\,\,\right\} \bigskip\\
 & \leq & \beta_{qr}^{\max}\left\{ \,\left({\max\limits _{k\,\in\mathcal{N}}\dfrac{\sigma_{r}^{2}(k)}{\sigma_{q}^{2}(k)}}\right)\,+\left[{\max\limits _{r^{\,\prime}\in\Omega}}\,\left({\max\limits _{k\,\in\mathcal{N}}}\,{\dfrac{\sigma_{r^{\,\prime}}^{2}(k)}{\sigma_{q}^{2}(k)}}\right)\right]\,\left({\dfrac{\rho}{\chi}}-1\,\right)\,\right\} ,\quad\forall q,r\in\Omega,\text{ and }q\neq r,\end{array}\label{eq:upbetahat}\end{equation}
 where $\beta_{qr}^{\max}$ is defined in (\ref{eq:def:beta_max_hat}).

From (\ref{eq:upbetahat}), one infers that condition (\ref{SF_2_Uniqueness})
implies that $\overline{\mathbf{B}}(\mathbf{R}^{\star})$ in (\ref{Beta_bar_matrix})
is diagonally dominant, which is sufficient to guarantee the P-property
of $\overline{\mathbf{B}}(\mathbf{R}^{\star})$ \cite[Theorem 6.2.3]{BPlemmons79},
and thus the uniqueness of the GNE of (\ref{Power Game}) (Theorem
\ref{th:uniqueness_(main body)}).

It remains to show that (\ref{eq:SF_uniq_equal_sigmas}) is equivalent
to (\ref{SF_2_Uniqueness}) if $\sigma_{q}^{2}(k)=\sigma_{r}^{2}(k),$
$\forall r,q\in\Omega$ and $k\in\mathcal{N}$. In this case, (\ref{SF_2_Uniqueness})
reduces to \begin{equation}
\rho\,{\sum_{r\neq q}}\,\beta_{qr}^{\max}<\,\chi\, e^{-2R_{q}^{\star}},\hspace{1.0266pc}\forall\, q\,\in\Omega,\end{equation}
 or equivalently \begin{equation}
\rho\,{\max_{q\,\in\Omega}}\,\left[(e^{R_{q}^{\star}}-1){\sum_{r\neq q}}\,\beta_{qr}^{\max}\,\right]\,<\,\chi\,{\max_{q\,\in\Omega}}(\, e^{-R_{q}^{\star}}-e^{-2R_{q}^{\star}}\,),\end{equation}
 which is clearly equivalent to (\ref{eq:SF_uniq_equal_sigmas}).
\hspace{\fill}\rule{1.5ex}{1.5ex}

\section{Proof of Theorem \ref{th:Convergence_IWFA} and Theorem \ref{th:Convergence_SIWFA}}

\label{proof_th:Convergence_IWFA-SIWFA} The proof of the convergence
of both the sequential and simultaneous IWFAs\ is similar to the
proof of the uniqueness of the GNE of the game ${\mathscr{G}}$ 
as given in Appendix \ref{proof_th:uniqueness_(main body)}. The difference
is that instead of working with two solutions of the GNEP 
(and showing that they are equal under certain conditions), we consider
the users' power allocation vectors produced by the algorithms in
two consecutive iterations and derive conditions under which their
respective distances to the unique solution of the game contract under
some norm.

We focus first on the convergence of the simultaneous IWFA. Then,
we briefly show that a similar analysis can be carried out also for
the sequential IWFA. Throughout the following analysis, we assume
that conditions of Theorem~\ref{th:uniqueness_(main body)} are satisfied.

\subsection{Convergence of simultaneous IWFA}

Let us define ${\mathbf{p}}^{\,(n+1)}\triangleq({\mathbf{p}}_{q}^{(n+1)})_{q=1}^{Q},$
with ${\mathbf{p}}_{q}^{(n+1)}\triangleq(p_{q}^{(n+1)}(k))_{k=1}^{N}$
denoting the power allocation vector of user $q$ at iteration $n+1$
of the simultaneous IWFA given in Algorithm $1,$ and let \begin{equation}
r_{q}^{(n+1)}(k)\,\triangleq\,\log\left(\,1+{\frac{|H_{qq}(k)|^{2}\, p_{q}^{(n+1)}(k)}{\tau_{q}^{(n)}(k)}}\,\right),\hspace{1.0499pc}\, k\in\mathcal{N},\text{ }q\in\Omega,\text{ }n=0,1,\ldots,\end{equation}
 with $\tau_{q}^{(n)}(k)\triangleq\tau_{q}(k,\mathbf{p}^{(n)}(k)),$
where $\mathbf{p}^{(n)}(k)\triangleq(p_{q}^{\,(n)}(k))_{q=1}^{Q}$
and $\tau_{q}(k,\mathbf{p}^{(n)}(k))$ is defined as in (\ref{eq:def_tau_i_p_k}).
According to the simultaneous IWFA, at iteration $n+1,$ the power
allocation ${\mathbf{p}}_{q}^{(n+1)}$ of each user $q$ must satisfy
the single-user waterfilling solution (\ref{WF_single-user}) (see
(\ref{SIWFA_op})), given the allocations ${\mathbf{p}}_{-q}^{\,(n)}\triangleq(\mathbf{p}_{r}^{(n)})_{r\neq q=1}^{N}$of
the other users at the previous iteration. It follows that each ${\mathbf{p}}_{q}^{(n+1)}$
and $r_{q}^{(n+1)}(k)$ satisfy (see (\ref{KKT_VI})) \begin{equation}
\begin{array}{lll}
0\,\leq\, r_{q}^{(n+1)}(k) & \perp & \log(\tau_{q}^{(n)}(k))+r_{q}^{(n+1)}(k)-\log(|H_{qq}(k)|^{2})+\nu_{q}^{(n+1)}\geq\,0,\hspace{1.0499pc}k\,\in\mathcal{N}\text{,}\\[5pt]
\nu_{q}^{(n+1)}\text{ free,} &  & {\displaystyle \sum\limits _{k=1}^{n}}\, r_{q}^{(n+1)}(k)=R_{q}^{\star},\end{array}\text{ \quad}\forall n=0,1,\ldots,\end{equation}
 or equivalently (see (\ref{eq:VI})), for all $q\in\Omega$ and $n=0,1,\ldots,$
\begin{equation}
{\sum_{k=1}^{N}}\,\left(\, r_{q}(k)-r_{q}^{(n+1)}(k)\,\right)\,\left(\,-\log(|H_{qq}(k)|^{2})+\log(\tau_{q}^{(n)}(k))+r_{q}^{(n+1)}(k)\,\right)\,\geq\,0,\quad\forall\mathbf{r}_{q}\,\in U_{q},\label{eq:VI_IWFA}\end{equation}
 where $U_{q}$ is defined in (\ref{eq:VI-def}). Let ${\mathbf{p}}^{\star}\triangleq(\mathbf{p}_{q}^{\star})_{q=1}^{Q}$
denote the unique GNE of the game ${\mathscr{G}}$ 
(i.e.,the unique solution of (\ref{eq:DSL achievable})) and ${\mathbf{r}}^{\star}\triangleq(\mathbf{r}_{q}^{\star})_{q=1}^{Q}$
the unique rate solution of (\ref{eq:DSL achievable in rates}). Note
that each $\mathbf{r}_{q}^{\star}$ satisfies the constraints in (\ref{eq:VI_IWFA}).
Hence, following the same approach as in Appendix \ref{proof_th:uniqueness_(main body)}
to obtain (\ref{eq:uie-1}) from (\ref{eq:VI}), we deduce \begin{equation}
\left\Vert \mathbf{r}_{q}^{(n+1)}-\mathbf{r}_{q}^{\star}\,\right\Vert _{2}\,\leq\,\left\Vert \,\log(\boldsymbol{\tau}_{q}^{(n)})-\log(\boldsymbol{\tau}_{q}^{\star})\,\right\Vert _{2},\text{\quad}\forall q\in\Omega,\text{ }\forall n=0,1,\ldots,\label{eq:rates_ineq_IWFA}\end{equation}
 where $\boldsymbol{\tau}_{q}^{\star}\triangleq(\tau_{q}^{\star}(k))_{k=1}^{N}$
and $\tau_{q}^{\star}(k)\triangleq\tau_{q}(k,\mathbf{p}^{\star}(k)).$
Similarly to (\ref{eq:Delta_r_k}), there exists some scalar $s_{q}^{(n)}(k)$
such that \begin{equation}
\frac{|H_{qq}(k)|^{2}p_{q}^{(n+1)}(k)}{\tau_{q}^{(n)}(k)}\leq s_{q}^{(n)}(k)\leq\frac{|H_{qq}(k)|^{2}p_{q}^{\star}(k)}{\tau_{q}^{\star}(k)},\end{equation}
 and, for all $\, k\in\mathcal{N}$,{ }$q\in\Omega,$ and $n=0,1,\ldots,$\begin{equation}
\begin{array}{l}
r_{q}^{(n+1)}(k)-r_{q}^{\star}(k)=\left(\,\left(\, p_{q}^{\,(n+1)}(k)-p_{q}^{\ast}(k)\,\right)+{\dfrac{p_{q}^{\,(n+1)}(k)}{\tau_{q}^{(n)}(k)}}\,{\displaystyle \sum_{r\neq q}}\,|H_{qr}(k)|^{2}\left(p_{r}^{\,(n+1)}(k)-p_{r}^{\star}(k)\,\right)\,\right)\\
\hspace{3.5cm}\cdot{\dfrac{|H_{qq}(k)|^{2}}{\tau_{q}^{\star}(k)(\,1+s_{q}^{(n)}(k)\,)}.}\end{array}\label{eq:delta_r_IWFA}\end{equation}
 Moreover, there exists some scalar $\omega_{q}^{\,(n)}(k)$ such
that \begin{equation}
\tau_{q}^{(n)}(k)\leq\omega_{q}^{\,(n)}(k)\leq\tau_{q}^{\star}(k),\end{equation}
 and, for all $\, k\in\mathcal{N}$,{ }$q\in\Omega,$ $n=0,1,\ldots,$
\begin{equation}
\log(\tau_{q}^{(n)}(k))-\log(\tau_{q}^{\ast}(k))\,=\,{\frac{1}{\omega_{q}^{\,(n)}(k)}}\left(\tau_{q}^{(n)}(k)-\tau_{q}^{\ast}(k)\,\right)\,={\frac{1}{\omega_{q}^{\,(n)}(k)}}\,{\sum_{r\neq q}}\,|H_{qr}(k)|^{2}\left(p_{r}^{(n)}(k)-p_{r}^{\ast}(k)\right).\label{eq:delta_log_tau_IWFA}\end{equation}
 Introducing the vector $\mathbf{t}^{(n)}\triangleq(\mathbf{t}_{q}^{(n)})_{q=1}^{Q},$
with $t_{q}^{(n)}$ defined as \begin{equation}
t_{q}^{(n)}\,\triangleq\,\sqrt{{\sum_{k=1}^{N}}\,\left(\,{\frac{|H_{qq}(k)|^{2}\,\left(p_{q}^{(n)}(k)-p_{q}^{\ast}(k)\,\right)}{\bar{\tau}_{q}(k)}}\,\right)^{2}},\hspace{1.0499pc}q\in\Omega,\text{ }n=0,1,\ldots.,\label{eq:def_error_vec_IWFA}\end{equation}
 with $\bar{\tau}_{q}(k)$ given in (\ref{eq:upper_bound_tau_i_p_k})
and using (\ref{eq:delta_r_IWFA}) and (\ref{eq:delta_log_tau_IWFA}),
(\ref{eq:rates_ineq_IWFA}) leads to \begin{equation}
\left(\mbox{Diag}\left\{ \mathbf{B}(\mathbf{R}^{\star})\right\} \right)\,\mathbf{t}^{(n+1)}\,\leq\,\left(\mbox{off-Diag}\left\{ \mathbf{B}(\mathbf{R}^{\star})\right\} \right)\mathbf{t}^{(n)},\quad\text{ }n=0,1,\ldots.,\label{eq:vector contracts}\end{equation}
 where $\mbox{Diag}\left\{ \mathbf{B}(\mathbf{R}^{\star})\right\} $
and $\mbox{off-Diag}\left\{ \mathbf{B}(\mathbf{R}^{\star})\right\} $
are the diagonal and the off-diagonal parts of $\mathbf{B}(\mathbf{R}^{\star}),$
respectively, with $\mathbf{B}(\mathbf{R}^{\star})$ defined in (\ref{eq:matrix B}).
Based on (\ref{eq:vector contracts}), the proof of convergence of
the simultaneous IWFA is guaranteed under conditions on Theorem \ref{th:uniqueness_(main body)},
as argued next.

According to\ \cite[Lemma 5.3.14]{CPStone92}, the P-property of
$\overline{\mathbf{B}}(\mathbf{R}^{\star}),$ with $\overline{\mathbf{B}}(\mathbf{R}^{\star})$
defined in (\ref{Beta_bar_matrix}), is equivalent to the spectral
condition:\begin{equation}
\rho\left[\left({\mbox{Diag}\left\{ \mathbf{B}(\mathbf{R}^{\star})\right\} }\right){^{-1}\mbox{off-Diag}}\left\{ \mathbf{B}(\mathbf{R}^{\star})\right\} \right]{<1,}\label{eq:spectrum-cond}\end{equation}
 where $\rho(\mathbf{A})$ denotes the spectral radius of $\mathbf{A}$.
Therefore, by (\ref{eq:vector contracts}) and (\ref{eq:spectrum-cond}),
the sequence $\{\mathbf{t}^{(n)}\}$ contracts under a certain matrix
norm; hence it converges to zero. The claimed convergence of the sequence
$\{{\mathbf{p}}^{(n)}\}$ follows readily. 
\hspace{\fill}\rule{1.5ex}{1.5ex}

\subsection{Convergence of sequential IWFA}

The convergence of the sequential IWFA described in Algorithm 2 can
be studied using the same approach as for the simultaneous IWFA. The
difference is the final relationship between the error vectors in
two consecutive iterations of the algorithm. More specifically, using
the vectors $\mathbf{t}^{(n)}\triangleq(\mathbf{t}_{q}^{(n)})_{q=1}^{Q},$
with $t_{q}^{(n)}$ defined as in (\ref{eq:def_error_vec_IWFA}),
one can see that the sequential IWFA leads to \begin{equation}
\left(\mbox{Diag}\left\{ \mathbf{B}(\mathbf{R}^{\star})\right\} -\mbox{Low}\left\{ \mathbf{B}(\mathbf{R}^{\star})\right\} \right)\mathbf{t}^{(n+1)}\,\leq\left(\mbox{Up}\left\{ \mathbf{B}(\mathbf{R}^{\star})\right\} \right)\,\mathbf{t}^{(n)},\quad\text{ }n=0,1,\ldots.,\label{eq:vector contracts_seq_IWFA}\end{equation}
 where $\mbox{Low}\left\{ \mathbf{B}(\mathbf{R}^{\star})\right\} $
and $\mbox{Up}\left\{ \mathbf{B}(\mathbf{R}^{\star})\right\} $ denotes
the strictly lower and strictly upper triangular parts of $\mathbf{B}(\mathbf{R}^{\star}),$
respectively. The above inequality implies\begin{equation}
\mathbf{t}^{(n+1)}\,\leq\left(\mbox{Diag}\left\{ \mathbf{B}(\mathbf{R}^{\star})\right\} -\mbox{Low}\left\{ \mathbf{B}(\mathbf{R}^{\star})\right\} \,\right)^{-1}\,\left(\mbox{Up}\left\{ \mathbf{B}(\mathbf{R}^{\star})\right\} \,\right)\mathbf{t}^{(n)}=\mathbf{\Upsilon t}^{(n)},\quad\text{ }n=0,1,\ldots.,\label{eq:vector contracts_seq_IWFA_2}\end{equation}
 where \begin{equation}
\mathbf{\Upsilon}\triangleq\left(\mbox{Diag}\left\{ \mathbf{B}(\mathbf{R}^{\star})\right\} -\mbox{Low}\left\{ \mathbf{B}(\mathbf{R}^{\star})\right\} \,\right)^{-1}\,\mbox{Up}\left\{ \mathbf{B}(\mathbf{R}^{\star})\right\} \,.\label{eq:def_gamma}\end{equation}
 In (\ref{eq:vector contracts_seq_IWFA}) we used the fact that, under
the P-property of the Z-matrix $\mbox{Diag}\left\{ \mathbf{B}(\mathbf{R}^{\star})\right\} -\mbox{Low}\left\{ \mathbf{B}(\mathbf{R}^{\star})\right\} $
(due to the fact that all its principal minors are less than one),
the inverse $\left(\mbox{Diag}\left\{ \mathbf{B}(\mathbf{R}^{\star})\right\} -\mbox{Low}\left\{ \mathbf{B}(\mathbf{R}^{\star})\right\} \,\right)^{-1}$
is well-defined and nonnegative entry-wise \cite[Theorem $3.11.10$]{CPStone92}.

According to (\ref{eq:vector contracts_seq_IWFA_2}), the convergence
of the sequential IWFA is guaranteed under the spectral condition\begin{equation}
\rho\left(\mathbf{\Upsilon}\right){<1,}\end{equation}
 which is equivalent to the P-property of $\overline{\mathbf{B}}(\mathbf{R}^{\star})$
\cite[Lemma 5.3.14]{CPStone92}, with $\overline{\mathbf{B}}(\mathbf{R}^{\star})$
defined in (\ref{Beta_bar_matrix}).
\hspace{\fill}\rule{1.5ex}{1.5ex}

\def\baselinestretch{1}

\end{document}